\documentclass[arxiv,usenatbib]{iupartex}
\usepackage{newtxtext,newtxmath}
\usepackage[T1]{fontenc}
\usepackage{ae,aecompl}

\usepackage{multicol}   
\usepackage{bm}		    
\usepackage{pdflscape}	

\title[Comprehensive Analysis of NGC 6793]{Comprehensive Analysis of Middle-Aged Open Cluster NGC 6793 in Vulpecula via Gaia DR3 Data}

\author[Ta\c {s}demir et al.]{%
S. Ta\c {s}demir$^{1\cc}$\orcid{0000-0003-1339-9148},
D. C. Çınar$^{1}$\orcid{0000-0001-7940-3731},
R. Canbay$^{2}$ \orcid{0000-0003-2575-9892},
S. Taştan$^{2}$\orcid{0009-0009-3372-3663},
W. H. Elsanhoury$^{3}$\orcid{0000-0002-2298-4026}, and Haroon, A. A.$^{4,5}$\orcid{0000-0002-8194-5836}
\affsep \\
$^1$Istanbul University, Institute of Graduate Studies in Science, Programme of Astronomy and Space Sciences, 34116, Beyaz{\i}t, Istanbul, Türkiye\\
$^2$Istanbul University, Faculty of Science, Department of Astronomy and Space Sciences, 34119, Beyaz\i t, Istanbul, Türkiye\\
$^3$Department of Physics, College of Science, Northern Border University, Arar, Saudi Arabia\\
$^4$Astronomy and Space Science Department, Faculty of Science, King Abdulaziz University, Jeddah, Saudi Arabia\\
$^5$Department of Astronomy, National Research Institute of Astronomy and Geophysics (NRIAG), 11421, Helwan, Cairo, Egypt}
\corres{S. Ta\c {s}demir}{tasdemir.seval@ogr.iu.edu.tr}

\pubyear{2025}

\doiheader{XXXXXXX/PAR.2025.00000}
\date{
	\pSubmit{20.01.2025} 
	\pRevReq{06.02.0025}
	\pLastRevRec{08.03.2025}
	\pAccept{21.03.2025}
	\pPubOnl{00.00.0000}
}
\volume{3}
\volnumber{1}

\begin{document}
\label{firstpage}
\pagerange{\pageref*{firstpage}--\pageref*{lastpage}}
\maketitle

\begin{abstract}
We conducted an in-depth analysis of NGC 6793 open cluster via \textit{Gaia} DR3 data, including astrometric, spectroscopic, and photometric measurements. Selection of 147 stars, which show membership probabilities $P\geq0.5$ were classified as likely members. The mean trigonometric parallaxes and proper-motion components of the cluster were found to be  $\varpi = 1.674 \pm0.045$ mas and ($\mu_{\alpha}\cos \delta$,~$\mu_{\delta}$) =($3.814\pm0.031$,~$3.547\pm0.034$) mas yr$^{-1}$. Fundamental astrophysical parameters of NGC 6793 are derived simultaneously as $t$ = $650 \pm 50$ Myr, $\mu$ = $9.508\pm0.070$ mag, and $E(G_{\rm BP}-G_{\rm RP})$ = $0.361 \pm 0.035$ mag. Additionally, the cluster's luminosity function analysis reveals the $G$-absolute magnitude limit of the most likely stars, indicating a well-defined stellar population. The total mass of the cluster, determined through the mass function (MF) and considering stars with membership probabilities $P \geq 0.5$, was estimated as 139 $\pm$ 12 $M/M_{\odot}$. The slope of the MF was found to be $\Gamma = 1.40 \pm 0.26$, a result consistent with the Salpeter value. The kinematic analyses present velocity ellipsoid parameters as well as the convergent point $(A_{\rm o},~D_{\rm o}) = \left(85^{\circ}.85 \pm 0^{\circ}.11,~3^{\circ}.12 \pm 0^{\circ}.57\right)$. Analyses have shown that it is moving in a box-shaped orbit beyond the Sun's galactic radius and belongs to the thin disk population of the Milky Way. The calculated relaxation time suggests that NGC 6793 has reached a dynamically relaxed state, where the dynamical evolution parameter $\tau$ significantly exceeds one. These results highlight both the cluster's internal stability and its connection to the thin disc population.
\end{abstract}


\section{Introduction}
\label{sec:Introduction}

Open clusters (OCs) are stellar assemblies characterised by their gravitationally bound nature, comprising stars that have formed together within the same molecular cloud. This process, resulting in the collapse and disintegration of giant molecular clouds, is a fundamental aspect of star formation in the Milky Way \citep{Lada_2003, Kroupa_2001}. Consequently, the member stars of an OC share a common origin, which implies similarities in their chemical compositions (e.g., metallicity), distances from the Sun, and formation timescales. These shared properties make OCs prime targets for studies involving stellar populations and Galactic structure.  

However, the masses of stars in open clusters vary considerably due to the complexity of the star forming phenomena. This mass variation is directly reflected in their broad range of luminosities, effective temperatures, and spectral types \citep{Bastian_2010}. Furthermore, as clusters evolve, dynamical interactions such as mass segregation and tidal stripping can modify their internal structure and membership, providing valuable insights into both stellar and dynamical evolution \citep{Binney_Tremaine_2008}.  

In addition to their utility in tracing the Galaxy’s formation history, OCs serve as benchmarks for calibrating stellar evolutionary models. Their relatively young ages, often less than a few hundred million years, enable astronomers to study the early phases of stellar evolution, while older clusters, like M67, offer constraints on the later phases of stellar lifecycles \citep{Perryman_1998}. These characteristics solidify OCs as essential laboratories for investigating a wide array of astrophysical phenomena, from initial mass functions to the chemical evolution of the Galactic interstellar cosmic gas.  

The information provided by several researchers in the literature for NGC 6793 is as follows: NGC 6793 ($\alpha=19^{\rm h} 23^{\rm m} 16^{\rm s}.08, \delta= +22^{\circ} 09^{\rm '} 32^{\rm''}3$; $l= 56^{\circ}.1860, b=+03^{\circ}.3180$, J2000) is located in the Vulpecula constellation. \citet{Cantat-Gaudin_2020} determined the astrophysics and astrometric parameters of more than 2,000 OCs via \textit{Gaia} DR2 catalogue. \citet{Cantat-Gaudin_2020} analysed proper-motion components of NGC 6793 based on astrometric and photometric selection criteria, 185 stars were found as probable members. Using these members, \citet{Cantat-Gaudin_2020} estimated the $V$-band extinction, distance modulus, distance, and cluster age as $A_{\rm V}=0.74$ mag, $(m-M)=8.99$ mag, 628 pc, and 309 Myr, respectively. \citet{Cantat-Gaudin_2020}, reported proper-motion components ($\mu_{\alpha}\cos\delta, \mu_{\delta})$ as ($3.79 \pm 0.18, 3.54 \pm 0.18$) mas yr$^{-1}$, identifying 185 probable members based on positional and kinematic criteria.

\begin{table*}
\setlength{\tabcolsep}{10pt}
\renewcommand{\arraystretch}{1}
\small
  \centering
  \caption{The astrophysical parameters of NGC 6793, as gathered from available literature, include key measurements: colour excess ($E(B-V)$), heliocentric distance ($d$), metallicity ([Fe/H]), age ($t$), mean proper-motion components ($\langle\mu_{\alpha}\cos\delta\rangle$, $\langle\mu_{\delta}\rangle$), and systemic radial velocity ($V_{\rm R}$).}
  \begin{tabular}{cccccccc}
    \hline
    \hline
$E(B-V)$ &  $d$ & [Fe/H] & $t$ &  $\langle\mu_{\alpha}\cos\delta\rangle$ &  $\langle\mu_{\delta}\rangle$ & $V_{\rm R}$ & Ref \\
(mag) &  (pc)  & (dex) & (Myr) & (mas yr$^{-1}$) & (mas yr$^{-1}$) & (km s$^{-1})$ &      \\
    \hline
0.331$\pm$0.119 & 586$\pm$0.53            & ---    &309 & 3.799$\pm$0.011 & 	3.535$\pm$0.012  & -20.02$\pm$0.76 &(01) \\      
0.238           & 628            & ---             &309                  & --- & ---   & --- &(02) \\  
---             & 618            & ---             &309                  & 3.789$\pm$0.08 & 3.544$\pm$0.09  & --- &(03) \\
---             & 618            & ---             &309                  & ---  & ---  & -$37\pm5.35$ &(04) \\
---             & 618            & ---             &309& 3.794$\pm$0.170 & 3.543$\pm$0.146   & ---  &(05) \\
0.326$\pm$0.023             & 589$\pm$7      & 0.234$\pm$0.076 &458$\pm$69           & 3.778$\pm$0.194 & 3.569$\pm$0.237   & -25.681$\pm$4.71 &(06) \\
0.238            &---             & --- & 650             & 3.795$\pm$0.186     & 3.544$\pm$0.177   & ---  &(07) \\
0.33$\pm$0.004          &610$\pm$40            &-0.09$\pm$0.10            &500$\pm$50           & 3.818$\pm$0.055 & 3.611$\pm$0.071   & --- &(08) \\  
0.218$\pm$0.006            & ---            & 0.00            &572$\pm$88           & --- & ---   & --- &(09) \\  
---             & ---            & 0.25            &66$\pm$4             & 3.774$\pm$0.232 & 3.577$\pm$0.257   & --- &(10) \\  
---             & ---            & ---             &---                  & --- & ---   & -16.59$\pm$1.45 &(11)\\
0.312           & 724            & ---             &495                  & --- & ---   & --- &(12) \\  

0.256$\pm$0.025 & 585$\pm$19 &0.10$\pm$0.05 &650$\pm$50& 3.814$\pm$0.031 & 3.547$\pm$0.034   & -20.87$\pm$0.67 &(13) \\
  \hline
    \end{tabular}%
    \\
(01) \citet{Hunt_2024},   
(02) \citet{Donada_2023},
(03) \citet{Tarricq_2022},
(04) \citet{Tarricq_2021},
(05) \citet{Poggio_2021},
(06) \citet{Dias_2021},
(07) \citet{Cantat-Gaudin_2020},
(08) \citet{Yontan_2019}
(09) \citet{Bossini_2019},
(10) \citet{Liu_2019},
(11) \citet{Soubiran_2018},
(12) \citet{Joshi_2016},
(13) This study %
\label{tab:Table-1}
  \end{table*}%

CCD $UBV$ analyses of the NGC 6793 via \textit{Gaia} DR2 astrometric data was performed by \citet{Yontan_2019}. From 87 member stars, they determined reddening, distance, metallicity as $E(B-V)$= 0.33$\pm$0.04 mag, $d$=610$\pm$40 pc, [Fe/H]=-0.09$\pm$0.10 dex, respectively. \citet{Yontan_2019} also derived the cluster age as 500$\pm$50 Myr.

\citet{Soubiran_2018} conducted a kinematic investigations of more than 800 OCs through \textit{Gaia} DR2 database. For NGC 6793, based on 13 stars, they estimated the radial velocity $V_{\rm R} = 24.32 \pm 0.73$ km s$^{-1}$ and the mean space-velocity components as ($U$, $V$, $W$) = (-20.45 $\pm$ 0.81, -5.91 $\pm$ 1.20, -5.66 $\pm$ 0.09) km s$^{-1}$. Using Bayesian statistical techniques and \textit{Gaia} DR2 data, \citet{Bossini_2019} calculated the age of NGC 6793 to be $t$ = $572^{+90}_{-160}$ Myr. This determination helps to the explaining of the cluster's evolutionary stage and its place in the broader context of OCs. Table \ref{tab:Table-1} shows the literature results of NGC 6793 to make comparison easier.

The outline of the paper is as follows: Section \ref{sec:Data} provides a detailed overview of the both astrometric and photometric datasets used for NGC 6793. In Section \ref{sec:Results}, the methodologies employed to derive the structural parameters of stars within the open cluster as well as focus on presenting and analysing the photometric membership probabilities and astrophysical parameters. Section \ref{sec_LF_and_MF} reflects the distribution of member stars in well-defined both luminosity and mass functions. Evolving times of NGC 6793 are computed in Section \ref{sec_dynamical} and their convergent point was drawn here in Section \ref{sec_AD}. Section \ref{sec_7} is devoted to Galactic orbital parameters. Finally, Section \ref{sec:Conclusion} offers a summary of the findings and conclusions.
 
\section{Data}
\label{sec:Data}
\subsection{Astrometric and Photometric Data}

\begin{figure}
\centering
\includegraphics[scale=0.60, angle=0]{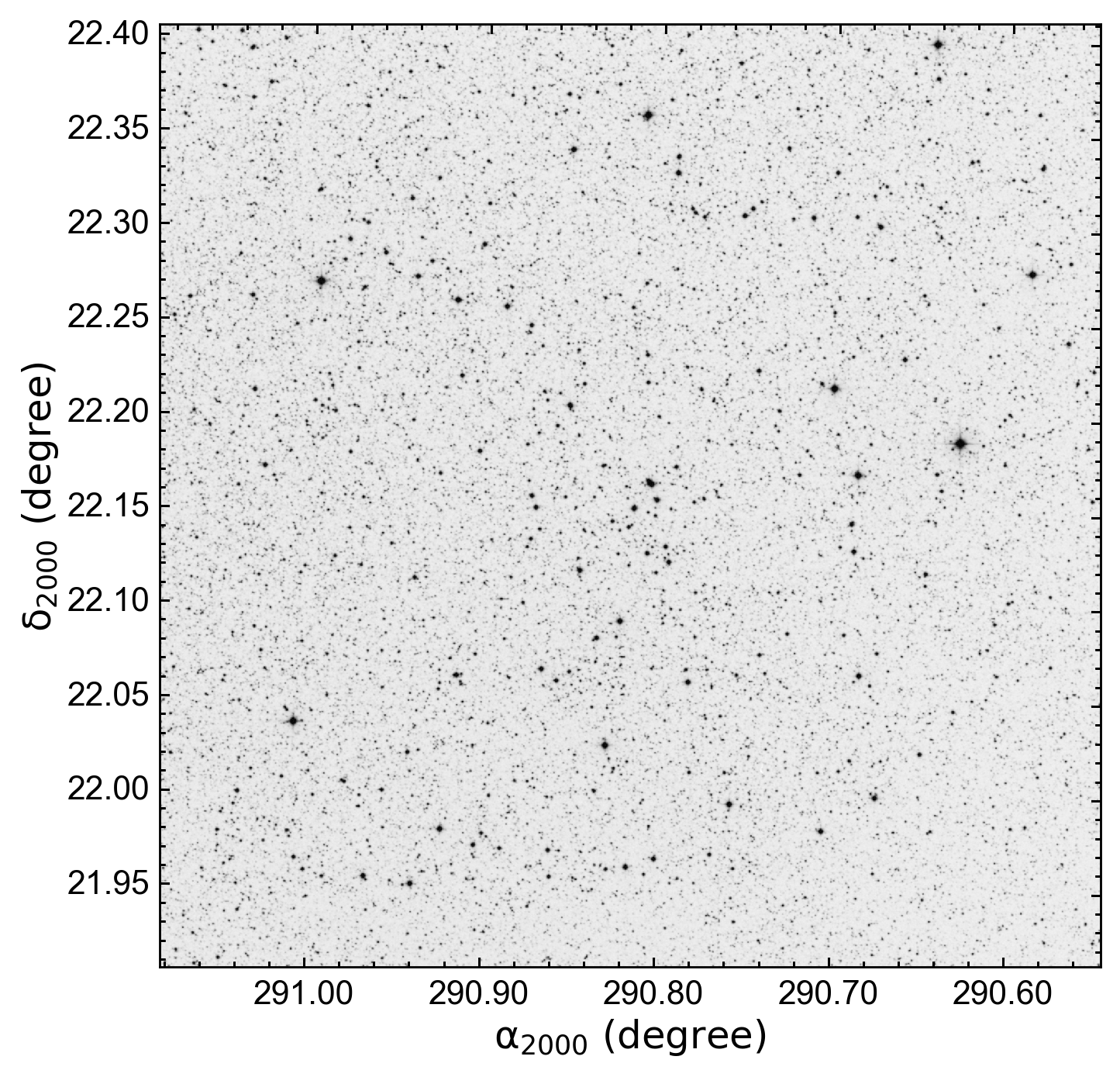}
\caption{The star chart of NGC 6793. The field of view is $30' \times 30'$.} 
\label{fig:ID_charts}
\end {figure}

With the help of the precise photometric (G, $G_{\rm BP}$ and $G_{\rm RP}$), astrometric ($\mu_{\alpha}\cos\delta, \mu_{\delta}$ and $\varpi$) and spectroscopic ($V_{\rm R}$) data from the latest data release of \textit{Gaia} \citep[\textit{Gaia} DR3,][]{Gaia_DR3}, we constructed an extensive catalogue for NGC 6793.  The equatorial coordinates were taken from the catalogue of \citet{Cantat-Gaudin-Anders_2020} are $\langle\alpha, \delta\rangle = (19^{\rm h} 23^{\rm m} 16^{\rm s}.08, \delta = +22^{\circ} 09^{\rm '} 32^{\rm''}.3)$, and we considered all stars within a 30 arcmin region centred around the cluster. The resulting catalogue includes 256,620 stars with $G$-apparent magnitudes $7 < G~{\rm (mag)}\leq 22$ interval. The star chart of the region around NGC 6793, obtained using the \texttt{POSS2UKSTU\_RED} filter\footnote{\url{https://archive.stsci.edu/cgi-bin/dss_form}}, is shown in Figure~\ref{fig:ID_charts}.

\vspace{0.8 cm}

\subsection{Photometric Completeness Limit}

\begin{figure}
\centering
\includegraphics[width=0.75\linewidth]{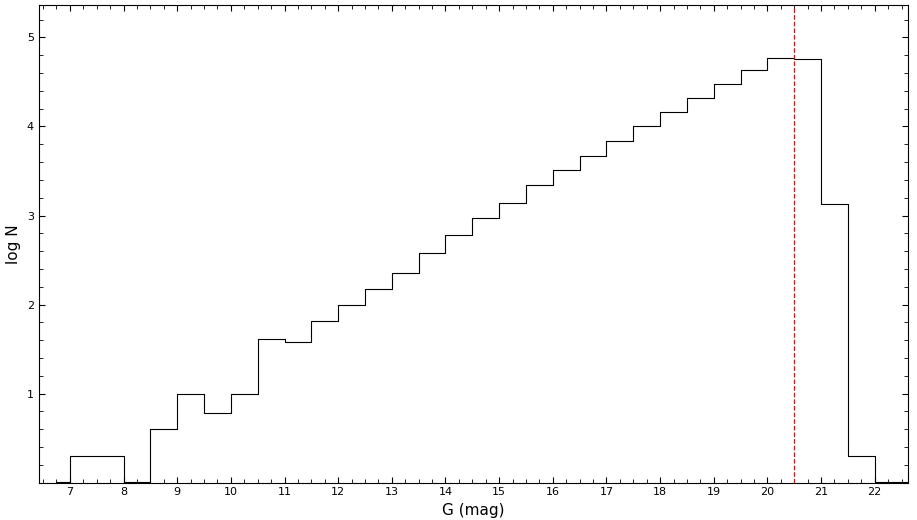}\\
\caption{$G$-apparent magnitude histogram of stars in the direction of the NGC 6793 OC. The dashed line represents the faint-limiting magnitude.}
\label{fig:histograms}
\end {figure}

In order to obtain accurate astrophysical and structural parameters of NGC 6793, the completeness limit of the cluster must first be established. This is crucial for accurate parameter estimation. In the case of NGC 6793, the photometric completeness limit was determined by analysing the star counts according to various $G$ magnitudes. As shown in Figure~\ref{fig:histograms}, the histogram reveals that the star count increases with brighter magnitudes up to a limit of $G=20.50$ mag, after which it decreases, indicating the completeness limit. Therefore, in subsequent analyses, stars fainter than this threshold were excluded statistically.

\begin{table}
\centering
\setlength{\tabcolsep}{10pt}
  \centering
  \caption{Errors in the apparent magnitudes ($G$) and colours ($G_{\rm BP}-G_{\rm RP}$) of stars in the cluster direction. }
    \begin{tabular}{cccc}
      \hline
  $G$ & $N$ & $\sigma_{\rm G}$ & $\sigma_{G_{\rm BP}-G_{\rm RP}}$\\
(mag) &  & (mag) & (mag) \\
\hline
(~7, 14]  & ~~~816  & 0.003   & 0.008   \\ 
(14, 15]  & ~1,258  & 0.003   & 0.007   \\ 
(15, 16]  & ~2,911  & 0.003   & 0.007   \\ 
(16, 17]  & ~6,521  & 0.003   & 0.009   \\ 
(17, 18]  & 13,904  & 0.003   & 0.016   \\ 
(18, 19]  & 28,137  & 0.003   & 0.034   \\ 
(19, 20]  & 55,495  & 0.005   & 0.078   \\ 
(20, 21]  & 95,811  & 0.010   & 0.200   \\ 
(21, 22]  & 16,193  & 0.028   & 0.426   \\ 
   \hline
    \end{tabular}%
  \label{tab:photometric_errors}%
\end{table}%

\vspace{1 cm}

\vspace{-0.95cm}
For error estimation, the uncertainties in the \textit{Gaia} DR3 data were treated as interval errors. The mean $G$ apparent magnitudes and $G_{\rm BP}-G_{\rm RP}$ colour indices of the stars in the open cluster field were calculated within $G$ magnitude intervals. At the completeness limit of $G = 20.50$ mag, the mean internal $G$-apparent magnitude and $G_{\rm BP}-G_{\rm RP}$ colour index were found to be 0.007 and 0.142 mag, respectively. The mean photometric errors of NGC 6793 as a function of apparent $G$ magnitudes are detailed in Table \ref{tab:photometric_errors}.

\vspace{-0.75cm}

\section{Method and Results}\label{sec:Results}
\subsection{Structural Parameters of NGC 6793}
To analyse the size and structure of NGC 6793, we performed a radial density profile (RDP) study. Taking into account the high-quality \textit{Gaia} DR3 data within a 30 arcminute radius, the cluster region was divided into several concentric rings based on the central coordinates from \citet{Cantat-Gaudin_2020}. The stellar density, $\rho(r)$, was determined by selecting stars to the limiting apparent $G=20.5$ mag.

\begin{figure}[]
\centering
\includegraphics[width=0.70\columnwidth]{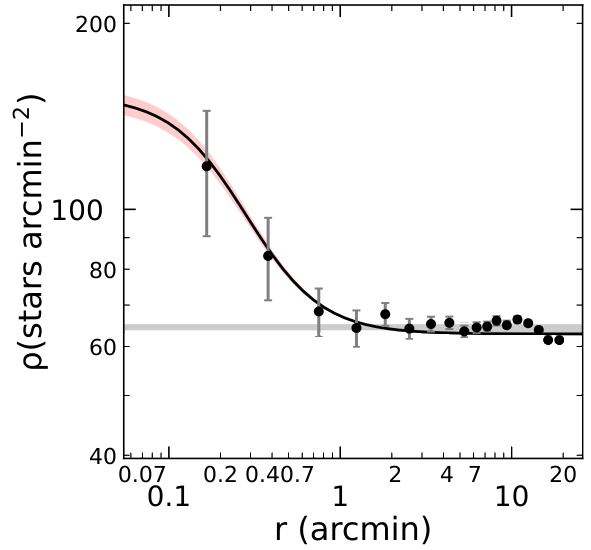}\\
\caption{Displayed is the radial density profile of NGC 6793. The black curve corresponds to the RDP profile described by \citet{King_1962}, with the horizontal grey band marking the stellar density of the background. The red-shaded region highlights the $1\sigma$ uncertainty in the King profile fit.} 
\label{fig:king}
\end {figure} 

For each ring, the stellar density was calculated using the formula $R_{i} = N_{i}/A_{i}$, where $N_{i}$ represents the number of stars in the $i^{\rm th}$ ring, and $A_{i}$ is the corresponding area of that ring. To assess the uncertainties, we applied the Poisson statistical error formula $1/\sqrt N$, where $N$ is the number of stars in the ring. Finally, the radial density profile was plotted, and the empirical model from \citet{King_1962} was used to fit the data:
\begin{equation}
\rho(r)=f_{\rm bg}+\frac{f_o}{1+(r/r_{\rm c})^2}
\end{equation}
In this context, the radius of the cluster is represented by $r$. The parameters $f_{\rm bg}$, $f_o$, and $r_{\rm c}$ refer to the stellar density of the background, the central stellar density, and the core radius, respectively. To derive the best estimates for these parameters, the $\chi^{2}$ minimization technique was employed during the RDP fitting process. The solid black line in Figure \ref{fig:king} demonstrates the best-fit solution to the radial density profile. The goodness of fit, indicated by the correlation coefficient $R^2=0.967$, suggests a high degree of accuracy in the estimation of the structural parameters, supporting the reliability of the model.

Based on the analysis of the radial density profile (RDP), we calculated the limiting radius of the cluster, $r_{\rm lim}^{\rm obs}$, to be $10$ arcmin. The derived structural parameters include $f_{\rm bg} = 62.855 \pm  0.750 2$, $f_{\rm 0} = 89.628 \pm 5.878$ stars arcmin$^{-2}$, and $r_{\rm c} = 1.593 \pm 0.128$ arcmin. To assess the reliability of the observed limiting radius ($r_{\rm lim}$), we applied the formula from \citet{Bukowiecki_2011}, which is as follows:
\begin{equation}
r_{\rm lim}=r_{\rm c}\sqrt{\frac{f_{\rm 0}}{3\sigma_{\rm bg}}-1}
\end{equation}
Using the formula, we determined the theoretical limiting radius to be 9.93 arcmin. This value aligns closely with the observed limiting radius $r_{\rm lim}^{\rm obs}$, indicating a strong consistency between the theoretical and measured parameters. 

\subsection{Membership Probabilities of Stars}

To ensure the distinction between open cluster stars and surrounding field stars, it is essential to account for contamination by both foreground and background stars, especially since OCs are often located in the Galactic plane. This contamination can hinder the accurate determination of astrophysical parameters, but it can be overcome by utilising proper-motion data, which reflect the common movement of cluster stars formed under similar conditions. The high precision of \textit{Gaia} DR3 data provides an essential tool for distinguishing these stars, as their vectorial motions are correlated. This makes proper-motion components a crucial factor in membership determination, ensuring that only the cluster stars are included in further analyses, as demonstrated in studies such as those by \citet{Bisht_2020}. The accuracy of the \textit{Gaia} DR3 data guarantees robust results in these membership studies, facilitating reliable astrophysical assessments.

In order to calculate the membership probabilities ($P$) of stars in NGC 6793, we employed the {\sc UPMASK} method, utilising the precise astrometric data from the \textit{Gaia} DR3 catalogue. This method is based on the $k$-means clustering technique, which identifies groups of stars with similar proper motions and parallaxes, providing a statistical probability for each star's membership in the cluster. Previous studies have applied this approach successfully \citep[i.e.][]{ Cantat-Gaudin_2020, Yontan_2022, Yontan_2023}. We performed 100 iterations of the method, using the astrometric parameters ($\alpha$, $\delta$, $\mu_{\alpha} \cos \delta$, $\mu_{\delta}$, $\varpi$) and their uncertainties. A total of 147 stars were identified as members, satisfying the criteria of $P \geq 0.5$, location within the limiting radius, and completeness limit.

\begin{figure*}
\centering
\includegraphics[width=1\textwidth]{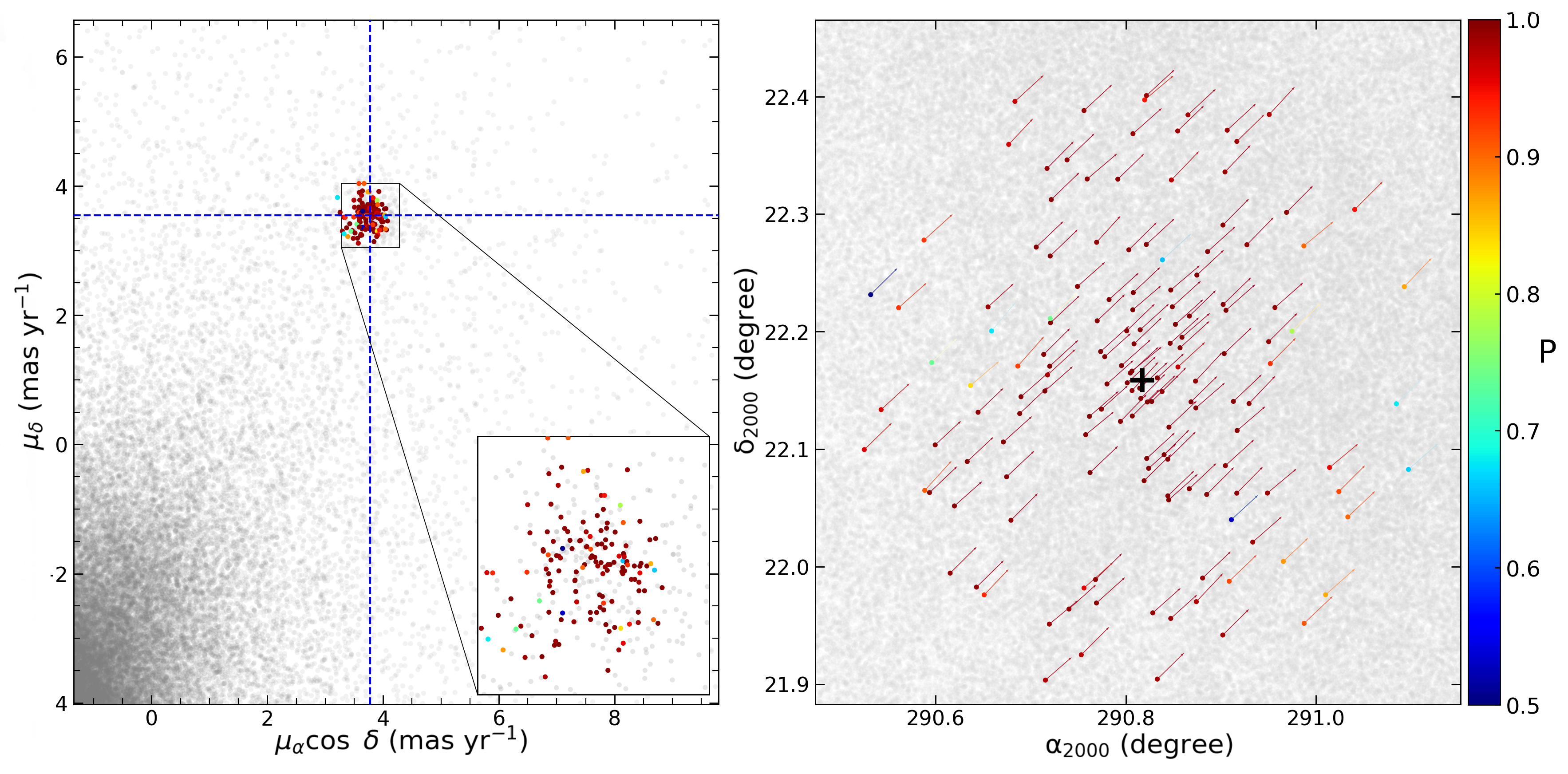}\\
\caption{Panel (a) presents the vector-point diagram (VPD), and panel (b) shows the proper-motion velocity vectors for NGC 6793 In the right panel, the colour scale represents most likely member stars ($P\geq0.5$). In panel (a), the magnified box highlights areas with a high density of member stars in the VPD, while the intersection of blue dashed lines indicates the mean proper motion values. In panel (b), the centre of NGC 6793's equatorial coordinates is marked with black crosshairs.
\label{fig:VPD_all}} 
 \end {figure*}

\begin{figure}
\centering
\includegraphics[width=0.5\textwidth]{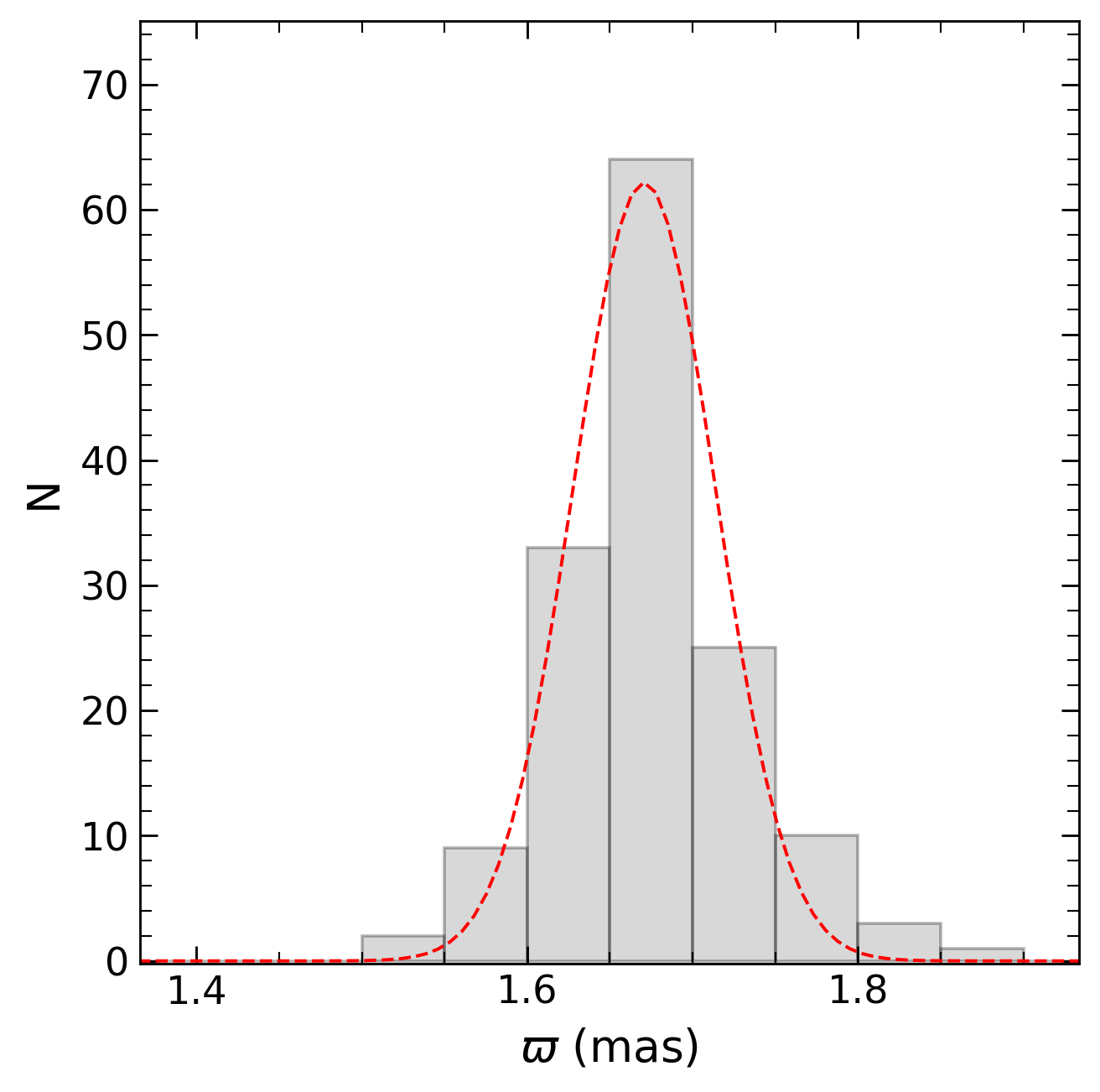}\\
\caption{Trigonometric parallax histogram of NGC 6793 from the member stars. The fitted Gaussian function is represented by the red dashed line.
\label{fig:plx_hist}}
\end {figure}
 
To identify the most likely stars of NGC 6793, we generated a vector-point diagram (VPD) by plotting the proper-motion components of the member stars. The VPD, shown in Figure ~\ref{fig:VPD_all}, clearly differentiates the NGC 6793 from the surrounding field stars. The blue dashed lines in the diagram represent the mean proper-motion components calculated from  $P\geq0.5$. This visualisation technique is commonly used in studies to isolate cluster members by their kinematic properties, offering a clear separation between cluster stars and foreground or background contamination.

The mean proper-motion components of NGC 6793 were obtained to be ($\mu_{\alpha}\cos \delta$, $\mu_{\delta}$) = (3.814$\pm$0.031,~3.547$\pm$0.034) mas yr$^{-1}$, reflecting the motion of the cluster's member stars. These measurements were derived from the most probable cluster members, which were selected based on their membership probabilities ($P \geq 0.5$). Additionally, we examined the trigonometric parallax data of the identified member stars and plotted the distribution using a histogram. By fitting Gaussian functions to the histogram, we calculated the mean trigonometric parallax value, as shown in Figure~\ref{fig:plx_hist}. The derived trigonometric parallax for NGC 6793 was $\varpi$=1.674$\pm 0.045$ mas, which provides a direct indication of the cluster's distance from the Sun.

To estimate the distance to NGC 6793, we applied the standard relation $d({\rm pc}) = 1000/\varpi$ (mas), which translates the parallax into a distance measurement in parsecs. Using this formula, we found the distance to be $d_{\varpi}$=597$\pm 27$ pc. This result is convenient with the findings of other researchers using \textit{Gaia} data, further confirming the accuracy and reliability of our measurement. A detailed comparison with other studies can be found in Table~\ref{tab:Table-1}, where the values derived for NGC 6793 in this work align well with those reported by various authors in the \textit{Gaia} era. This consistency underscores the robustness of the \textit{Gaia} astrometric data in determining the fundamental parameters of star clusters.

\subsection{Astrophysical Parameters of NGC 6793}

In determining fundamental astrophysical parameters, colour-magnitude diagrams (CMDs) serve as crucial instruments. The utilisation of these diagrams facilitates the identification of key features of a cluster, such as the main sequence, turn-off point, and giant stars, which are essential for deriving accurate cluster properties. In this study, we used the CMD derived from the \textit{Gaia} data for the most likely cluster members ($P\geq0.5$). In order to estimate the age, distance modulus, reddening, and metallicity of NGC 6793, we used the {\sc PARSEC} stellar isochrones \citep{Bressan_2012}. These isochrones, which are specifically designed for the \textit{Gaia} Early Data Release 3 \citep[{\it Gaia} EDR3;][]{Gaia_EDR3} photometric bands \citep{Riello_2021}, were fitted to the observed CMD of the cluster.

To derive the fundamental parameters of NGC 6793, we utilised the $G \times (G_{\rm BP} - G_{\rm RP})$ CMD. Isochrones were fitted through visual comparison, focusing on the most likely members ($P\geq 0.5$) that were identified as part of the main sequence, turn-off point, and giant stars. Before fitting the isochrones, we applied scaling based on the mass fraction $z$. This procedure provided the \textit{Gaia}-based colour excess of $E(G_{\rm BP} - G_{\rm RP}) = 0.361 \pm 0.035$ mag. To align our results with previous studies, we converted this value to the $UBV$-based colour excess, $E(B-V)$, using the relation $E(G_{\rm BP} - G_{\rm RP}) = 1.41 \times E(B-V)$ as suggested by \citet{Sun_2021}. Our resulting value of $E(B-V) = 0.256 \pm 0.025$ mag is consistent with the findings of various studies, confirming agreement across different analyses (see Table~\ref{tab:Table-1}).

We converted the estimated metallicity [Fe/H] = $0.10\pm 0.05$ dex to the mass fraction $z$ in order to select of isochrones as well as for the deriving of fundamental astrophysical parameters. To be able to transform, It is used an analytic expressions of Bovy\footnote{\url{https://github.com/jobovy/isodist/blob/master/isodist/Isochrone.py}} appropriate for {\sc parsec} isochrones \citep{Bressan_2012, Yontan_2023,Gökmen_2023}. The expressions are given as follow:
\begin{equation}
z_{\rm x}={10^{{\rm [Fe/H]}+\log \left(\frac{z_{\odot}}{1-0.248-2.78\times z_{\odot}}\right)}}
\end{equation}      
and
\begin{equation}
z=\frac{(z_{\rm x}-0.2485\times z_{\rm x})}{(2.78\times z_{\rm x}+1)}.
\end{equation} 
here $z_{\rm x}$ and $z_{\odot}$ represent intermediate values, with solar metallicity $z_{\odot}$ adopted as 0.0152 \citep{Bressan_2012}. For NGC 6793, we determined a metallicity of $z=0.0098$.

\begin{figure*}
\centering
\includegraphics[width=0.5\textwidth]{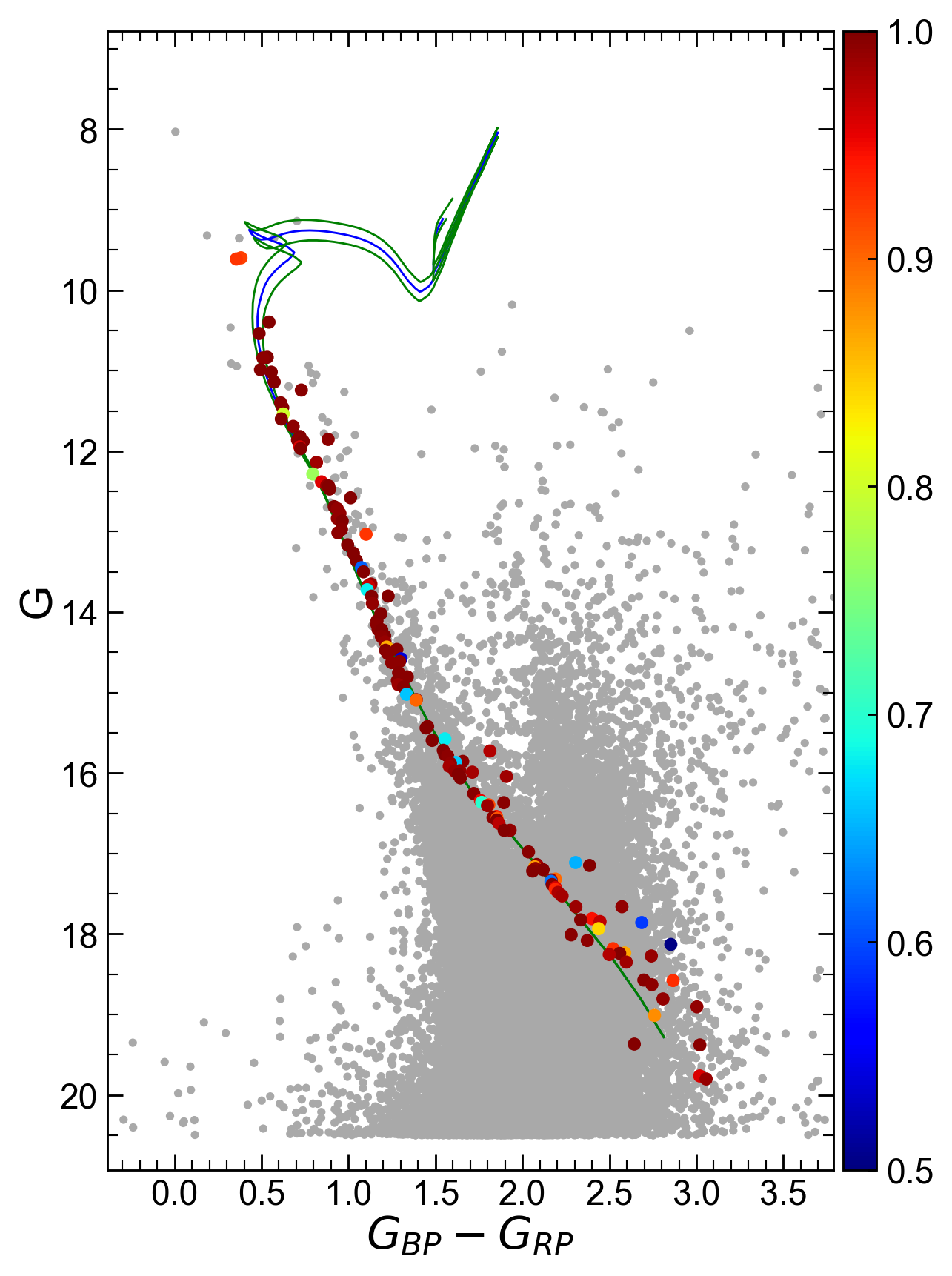}
\caption{CMD of NGC 6793 is displayed, with different colours representing the member stars of the cluster. The corresponding membership scales are indicated by the colour bars on the right. Stars with probabilities $P<0.5$ are marked as grey dots. The best-fit {\sc parsec} isochrones are shown in blue, while their uncertainties are illustrated with green lines. The overlaid isochrones correspond to an age of $650\pm50$ Myr for NGC 6793.
\label{fig:figure_age} }
\end {figure*}

The {\sc PARSEC} isochrones with ages of $\log~t = 8.77,~8.81$, and $8.85$ and metallicity $z=0.0098$ were fitted to the observed $G\times (G_{\rm BP}-G_{\rm RP})$ CMD for NGC 6793, as illustrated in Figure~\ref{fig:figure_age}. The best match was achieved for $\log~t = 8.81$, indicating an age of $t=650\pm 50$ Myr for the cluster. This age estimate aligns well with the results reported by \citet{Bossini_2019}, \citet{Zhong_2020}, and \citet{Cantat-Gaudin_2020} within their quoted uncertainties (see Table~\ref{tab:Table-1}). The derived distance modulus, $\mu=9.508\pm 0.070$ mag, corresponds to an isochrone distance of $d_{\rm iso}=585\pm 19$ pc. The uncertainties in the distance modulus and isochrone distance were estimated using the formulas given by \citet{Carraro_2017}, which incorporate the uncertainties in photometric magnitudes and excess values of the colour. 

The calculated isochrone distance is consistent with the trigonometric parallax distance of $d_{\varpi}=597\pm 26$ pc obtained in this study, as well as with the values of \citet{Yontan_2019} (see Table~\ref{tab:Table-1}). Additionally, the Galactocentric coordinates $(X, Y, Z)_{\odot}$ of NGC 6793 were determined. Here, $X$ represents the direction towards the Galactic centre, $Y$ the direction of Galactic rotation, and $Z$ the direction towards the Galactic north pole. Using the isochrone distance along with Galactic coordinates, these coordinates were calculated as $(X, Y, Z)_{\odot}=(325\pm18, 485\pm22, 34\pm6)$ pc. There is strong agreement between these values and those reported by \citet{Cantat-Gaudin_2020}.

\section{Luminosity and Mass Functions}
\label{sec_LF_and_MF}
\begin{figure}
    \centering
   \includegraphics[width=0.5\columnwidth]{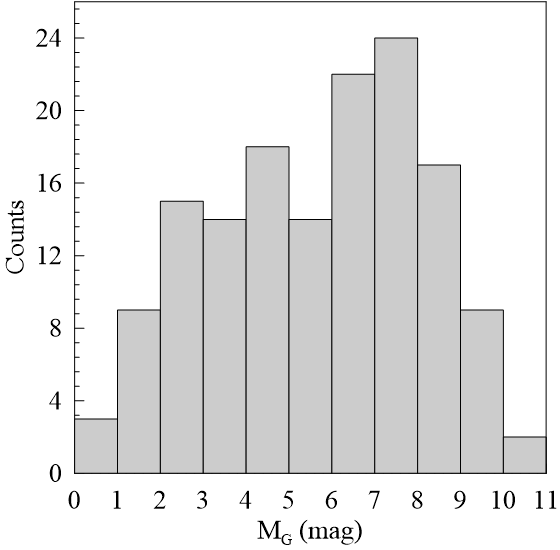}
    \caption{Histograms displaying the distribution of stars along the main sequence within each absolute $M_{\rm G}$ magnitude bin for NGC 6793.}
    \label{figure:LF}
\end{figure}

The luminosity function (LF) represents how stars within the cluster are distributed based on their brightness levels, while the mass function (MF) describes the distribution of stars according to their masses. Since all members of an open cluster (OC) originate simultaneously from the same molecular cloud under comparable physical conditions, these clusters serve as ideal subjects for analyzing stellar distributions in terms of magnitudes and masses. The LF for these cluster members is depicted in Figure~\ref{figure:LF}, and the corresponding mean absolute magnitudes ($\overline{M}_{\rm G}$) are provided in Table \ref{table:LF_MF}. The LF exhibits a modest upward trend, indicating that the cluster retains its low to intermediate-mass stars—a characteristic that is rather uncommon for older open clusters in the Milky Way. Based on calculations from the earlier section, the NGC 6793 OC resides in the inner Galactic disc at a Galactocentric distance of 7.69 $\pm$ 0.01 kpc, where it experiences significant tidal forces from this region.
A connection exists between LF, MF, and the mass-luminosity relation (MLR). For this study, the relation between absolute magnitudes $M_{\rm G}$ and stellar masses $M/M_\odot$ provided by \citet{Evans_2018} was utilized. This was done for a metallicity of $Z = 0.0098$ and $\log t=8.81$, covering absolute magnitudes within the range $0.08 < M_{\rm G}~{\rm (mag)}\leq 10.3$.

Once this relationship is established, the total mass of stars and star clusters within a specified magnitude range can be determined by substituting the observed $M_G$ values into the corresponding polynomial equation. This approach facilitates the estimation of stellar masses in cases where direct measurements are not feasible, leveraging the accurately calibrated MLR derived from isochrone models. The resulting values are summarized in Table \ref{table:LF_MF}. 
\begin{table}
\centering
\label{table:LF_MF}
\caption {Numerical values for the LF, MF, and the fitting parameters, including the slope of the MF for NGC 6793.}
\begin{tabular}{l|l|l}
\hline
\hline
Parameter    & Value & References  \\
\hline
$\overline{M}_{\rm G}$ (mag) & 5.66 $\pm$ 1.09  & This study    \\
$M_{ \rm C}$ ($M_{\odot}$) & 139 $\pm$ 12    & This study \\
$M_{\rm C}$ ($M_{\odot}$)& 625.45 $\pm$ 74.95  & \citet{Hunt_2024}   \\
$\overline{M}_{\rm C}$ ($M_{\odot}$) & 0.949  & This study    \\
$\Gamma$ & 1.40 $\pm$ 0.26  & This study    \\
\hline
\end{tabular}
\label{table:LF_MF}
\end{table}

\vspace{1cm}

\vspace{1cm}

\begin{figure}
    \centering
   \includegraphics[width=0.6\columnwidth]{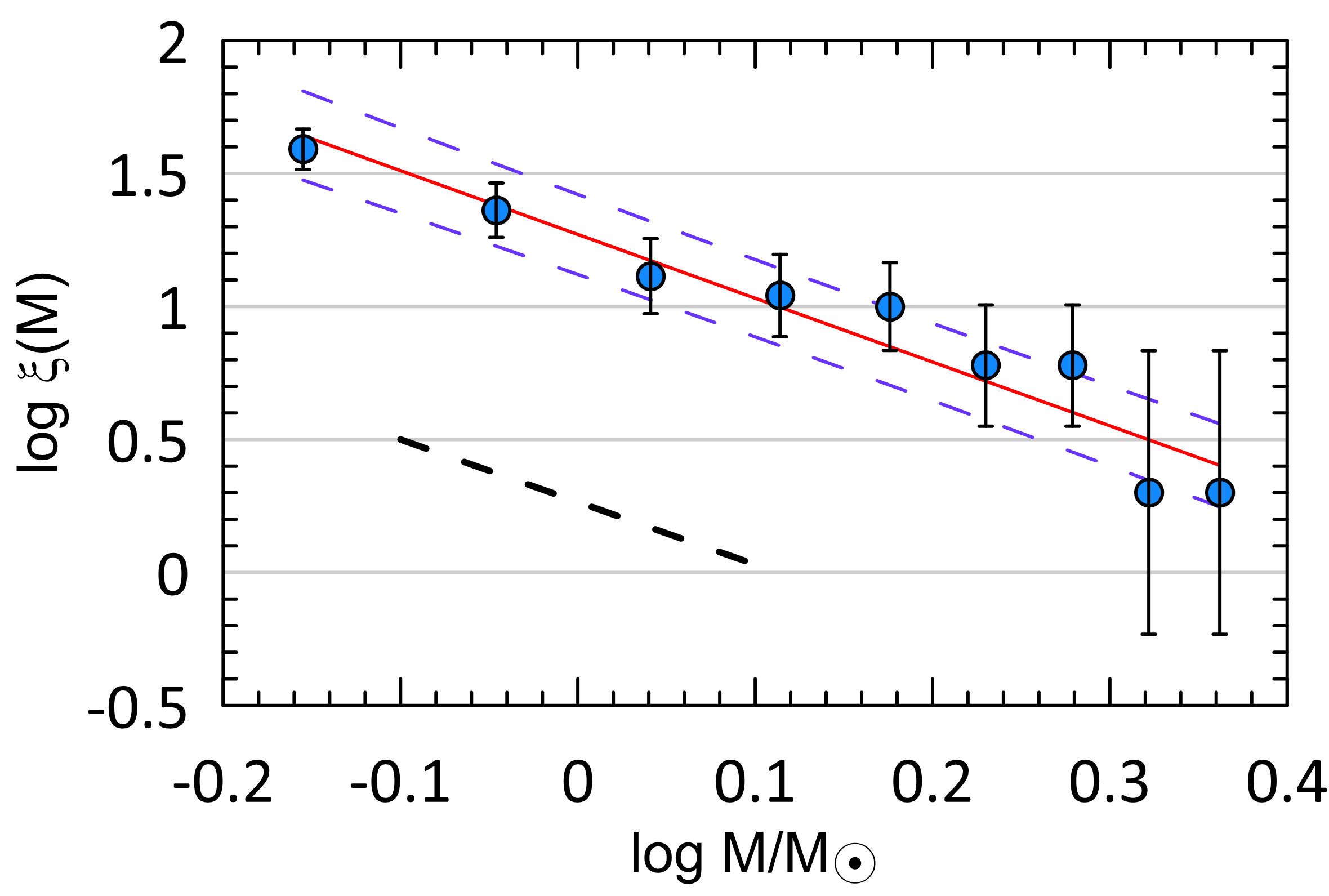}
       \caption{The cluster member mass function, fitted by the power-law of \citet{Salpeter_1955} (dashed black line), used to determine the slope ($\alpha$) for the clusters under investigation.}
    \label{figure:MLR}
\end{figure}

The mass spectra of open clusters span a wide range of stellar masses, encompassing both very low and high-mass stars. This diversity makes open clusters valuable for investigating the initial mass function (IMF), which represents the primary distribution of stellar masses. The IMF has been extensively explored in previous studies \citep{Phelps_janes_1993, Piskunov_2004, Yontan_2023, Gökmen_2023}.

\citet{Salpeter_1955} introduced the IMF as a measure of the number of stars ($dN$) distributed across a logarithmic mass scale ($dM$) for a given mass $M$. It is mathematically defined as $dN/dM = M^{-\alpha}$, where $\alpha = (1 + \Gamma)$. The value of $\alpha$ is 2.35, and the present-day mass function (MF) can be expressed as:
\begin{equation}
\label{eq:mf}
\log \left(\frac{dN}{dM}\right) = -(1+\Gamma) \times \log(M) + \text{constant}
\end{equation}
Here, $\alpha$ is a dimensionless parameter representing the slope of the MF, which is illustrated as a straight line in Figure~\ref{figure:MLR}. This slope, especially for massive stars ($M > 1~M_{\odot}$), acts as a marker for dynamical evolution. Salpeter’s power law indicates a rapid decline in the number of stars as mass increases.

In the study, the mass range considered for MF calculations is $0.08 < M_{\rm G}~{\rm (mag)} \leq 10.3$, encompassing a representative sample of stars. By utilising a least-squares fitting to the MF data, we derived a slope of $\Gamma = 1.40 \pm 0.26$, consistent with Salpeter's value within the margin of uncertainty. Furthermore, the total mass of the cluster ($M_{\rm C}$) and the mean stellar mass ($\overline{M/M_{\odot}}$) for NGC 6793 were calculated as $139 \pm 12~M_{\odot}$ and $0.949~M_{\odot}$, respectively. These findings are summarized in Table \ref{table:LF_MF}.

\section{Dynamical Relaxation and evolution times}
\label{sec_dynamical}
From a dynamical perspective, the interactions among stars in OCs lead to energy redistribution, distinguishing them from their compact halo counterparts such as globular clusters \citep{Inagaki_1985, Baumgardt_2003}. OCs exhibit a less dense spatial distribution, and processes such as contraction and disruption result in mass segregation, where massive stars concentrate closer to the cluster core compared to their less massive counterparts. This phenomenon has been observed in numerous OCs \citep{Dib_2018, Bisht_2020}. 

Regarding their kinetic energy, the velocity distribution of stars in OCs gravitates towards Maxwellian equilibrium \citep{Yadav_2013, Bisht_2019}. The time required for this dynamical adjustment is termed the dynamical relaxation time ($T_{\text{relax}}$), during which low-mass stars, possessing greater random velocities, occupy a larger spatial volume compared to massive stars \citep{Mathieu_1986}. The mathematical representation of $T_{\text{relax}}$, which depends on the number of cluster members ($N$) and the cluster diameter ($D \approx 2r_{\rm lim}$), is given as follows \citep{Spitzer_1971, Lada_2003, Maciejewski_2007}:
\begin{equation}
\label{eq:Trelax}
T_{\rm relax} = \frac{8.9 \times 10^5 \times N^{1/2} \times R_{\rm h}^{3/2}}{\overline{M}_{\rm C}^{1/2} \times \log(0.4 \times N)},
\end{equation}
where $\overline{M}_{\rm C}$ represents the mean stellar mass (in $M_\odot$) and $R_{\rm h}$ (in pc) is the radius enclosing approximately 50\% of the cluster mass. The latter can be derived using the relation \citep{Sableviciute_2006}:
\begin{equation}
R_{\rm h} = 0.547 \times r_{\rm c} \times \Big(\frac{r_{\rm t}}{r_{\rm c}}\Big)^{0.486},
\end{equation}
with $r_{\rm c}$ and $r_{\rm t}$ denoting the core and tidal radii, respectively. 

Alternatively, $T_{\rm relax}$ can be estimated using the dynamical crossing time ($T_{\rm cross} = D / \sigma_{\rm v}$), where $\sigma_{\rm v} = \sqrt{\sigma_1^2 + \sigma_2^2 + \sigma_3^2}$ is the velocity dispersion \citep{Lada_2003}. The crossing time $T_{\rm cross}$, defined as the time required for a cluster to complete one passage through the Galaxy, is approximately $10^6$ years and is independent of the cluster’s size or orbital shape \citep{Binney_1998}. 

In the Galactic disc, the gravitational influence on massive bodies like star clusters can be expressed using the formula from \citet{Röser_2019}:
\begin{equation}
x_{\rm L} = \left( \frac{G M_{\rm C}}{4A(A - B)} \right)^{1/3} = \left( \frac{G M_{\rm C}}{4\Omega_{o}^2 - \kappa^2} \right)^{1/3},
\end{equation}
where $x_{\rm L}$ approximates the cluster’s tidal radius ($r_{\rm t}$), and $G = 4.30 \times 10^{-6}$ kpc $M_\odot^{-1}$ (km s$^{-1}$)$^2$ is the gravitational constant. The parameters $\Omega_{\rm o} = A - B$ and $\kappa = \sqrt{-4B(A-B)}$ represent the angular velocity and the epicyclic frequency at the Sun’s position, respectively, with Oort’s constants $A$ and $B$ being $15.3 \pm 0.4$ km s$^{-1}$ kpc$^{-1}$ and $-11.9 \pm 0.4$ km s$^{-1}$ kpc$^{-1}$ \citep{Bovy_2015} or $9.38 \pm 0.33$ km s$^{-1}$ kpc$^{-1}$ and $-16.69 \pm 0.25$ km s$^{-1}$ kpc$^{-1}$ \citep{Elsanhoury_2024}. 
\begin{table}
\label{table_dynamical}
\centering
\caption {Dynamical evolution parameters for NGC 6793, including various times and the calculated escape velocity.}
\begin{tabular}{l|l}
\hline
\hline
Parameter    & Value \\
\hline
$T_{\rm relax}$ (Myr) & 4.26     \\
$\tau_{\rm ev}$ (Myr) & 426      \\
$\tau$ & 153       \\
$\sigma_{\rm 1}$ (km s$^{-1}$)& $25.43\pm5.04$\\
$\sigma_{\rm 2}$ (km s$^{-1}$)& $0.57\pm0.01$\\
$\sigma_{\rm 3}$ (km s$^{-1}$)& $0.51\pm0.01$\\
$\sigma_{\rm o}$ (km s$^{-1}$)& $25.44\pm5.04$\\
$T_{\rm cross}$ (Myr) & 0.136     \\
$V_{\rm esc}$ (km s$^{-1}$) & 222 $\pm$ 14.91  \\
\hline
\end{tabular}
\label{table_dynamical}
\end{table}

Based on our analysis, we estimate $r_{\rm t} \approx 7.11 \pm 0.38$ pc for NGC 6793. Consequently, the half-mass radius ($R_{\rm h}$) is determined to be $0.773 \pm 0.01$ pc, while the dynamical relaxation time ($T_{\rm relax}$) is found to be approximately 4.26 Myr.

The dynamical state of the cluster can be characterized by calculating its dynamical evolution parameter, expressed as $\tau = \text{age} / T_{\rm relax}$. For NGC 6793 OC, $\tau$ is found to be significantly greater than 1 ($\tau \gg 1$), indicating that the cluster has reached a state of dynamical relaxation.
Subsequently, our attention shifts to estimating the evaporation time ($\tau_{\rm ev} \simeq 10^2 T_{\rm relax}$), which represents the duration required for internal stellar interactions to completely remove all member stars \citep{Adams_2001}. Stars with lower masses continue to escape from the system, primarily at lower velocities, exiting through the Lagrange points \citep{Küpper_2008}. To maintain the cluster's gravitational binding, the escape velocity ($V_{\rm esc}$) in the context of rapid gas expulsion is defined as $V_{\rm esc} = R_{\rm gc} \sqrt{2GM_{\rm C}/3r_{\rm t}^3}$ \citep{Fich_1991, Fukushige_2000}. A detailed summary of the derived dynamical parameters, along with the corresponding timescales and escape velocity, is provided in Table \ref{table_dynamical}.

\section{The Convergent Point}
\label{sec_AD}
\begin{figure*}
    \centering
   \includegraphics[width=1\textwidth]{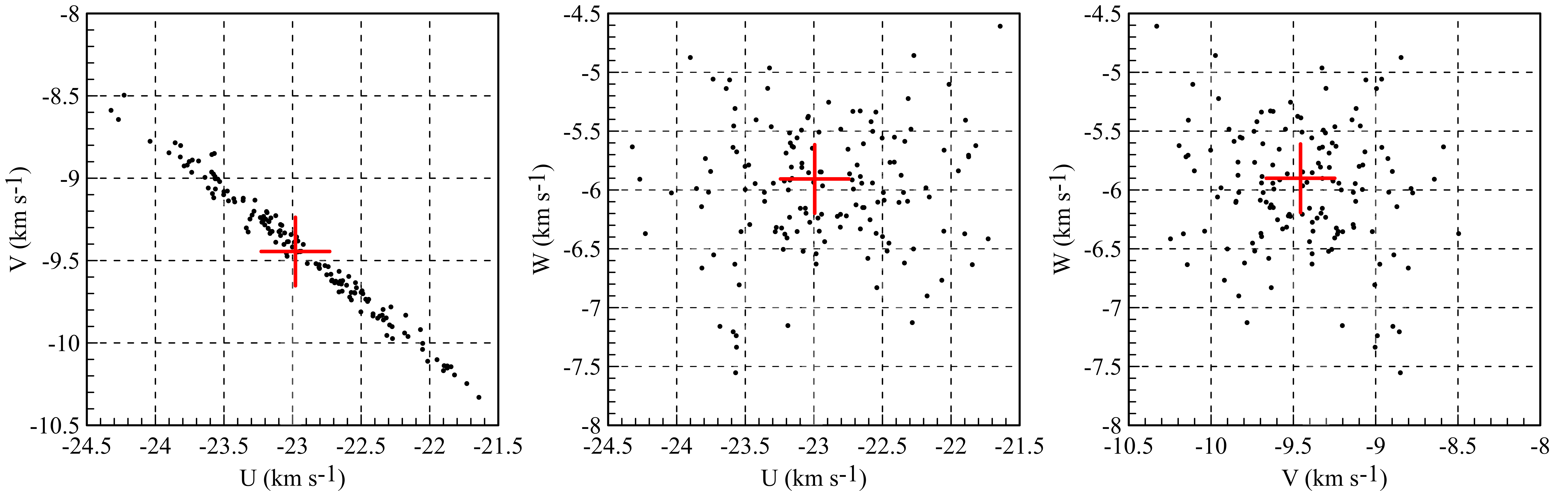}
    \caption{Distribution of the components of space velocity for NGC 6793.}
    \label{figure:uvw}
\end{figure*}

To examine the gravitational binding of stellar groups confined within a specific spatial region of the Galactic system, where their motions exhibit parallelism and uniformity, we assessed the velocity ellipsoid parameters (VEPs) and their kinematics using computational methods outlined by \citet{Bisht_2020} and \citet{Elsanhoury_2018}. 
The space velocity vectors ($V_{\rm x}$, $V_{\rm y}$, $V_{\rm z}$) for cluster members are calculated based on their coordinates ($\alpha$, $\delta$), distance ($d$), proper motion components ($\mu_{\alpha}\cos\delta$, $\mu_\delta$), and mean radial velocity ($V_{\rm R}$), using the methodology outlined by \citet{Melchior1958}:
\begin{equation}
\small
\label{eq:Vs}
\begin{pmatrix}
V_x \\
V_y \\
V_z
\end{pmatrix} 
= 
\begin{pmatrix}
-4.74~d~\mu_{\alpha}\cos{\delta}\sin{\alpha} - 4.74~d~\mu_{\delta}\sin{\delta}\cos{\alpha}  + V_{\rm R} \cos\delta \cos\alpha \\
+4.74~d~\mu_{\alpha}\cos{\delta}\sin{\alpha} - 4.74~d~\mu_{\delta}\sin{\delta}\cos{\alpha} + V_{\rm R} \cos\delta \cos\alpha \\
+4.74~d~\mu_\delta\cos\delta + V_{\rm R}\sin\delta
\end{pmatrix}
\end{equation}
Our calculations yielded a mean radial velocity for NGC 6793 as $V_{\rm R}=-20.87\pm 0.67$ km s$^{-1}$. Figure \ref{figure:uvw} illustrates the spatial velocity distribution of the stars most likely associated with NGC 6793. Utilising the computed space velocity components from Equation (\ref{eq:Vs}), we derived the Galactic velocity components ($U$, $V$, $W$) using the equatorial-to-Galactic transformation matrix obtained from the \textsc{SPECFIND} v2.0 radio continuum spectra catalogue \citep{Liu11}, expressed as:
\begin{equation}
\small
\label{eq:uvw}
\begin{pmatrix}
U \\
V \\
W
\end{pmatrix} 
= 
\begin{pmatrix}
-0.0518807421 \; V_{x} - 0.872222642 \; V_{y} - 0.4863497200 \; V_{z}\\
+0.4846922369 \; V_{x} - 0.4477920852 \; V_{y} + 0.7513692061 \; V_{z}\\
-0.873144899 \; V_{x} - 0.196748341 \; V_{y} + 0.4459913295 \; V_{z}\\
\end{pmatrix}
\end{equation}

Utilising the solar motion space velocity components provided by \citet{Coskunoglu_2011}, $(U, V, W){\odot} = (8.83\pm 0.24, 14.19\pm 0.34, 6.57\pm 0.21$) km s$^{-1}$, we adjusted the space velocity values $(U, V, W)$ of NGC 6793 to the Local Standard of Rest (LSR). The corrected values were found to be $(U, V, W)_{\rm LSR} = (-14.11\pm 0.25, 4.75\pm 0.97, 0.67\pm 0.26)$ km s$^{-1}$. Based on these corrected components, the total space velocity was calculated as $S_{\rm LSR} = 14.90\pm 1.04$ km s$^{-1}$. This velocity aligns well with the range expected for stars belonging to the young thin-disc population \citep{Leggett_1992}.

\begin{figure}
\begin{center}
\includegraphics[width=0.65\columnwidth]{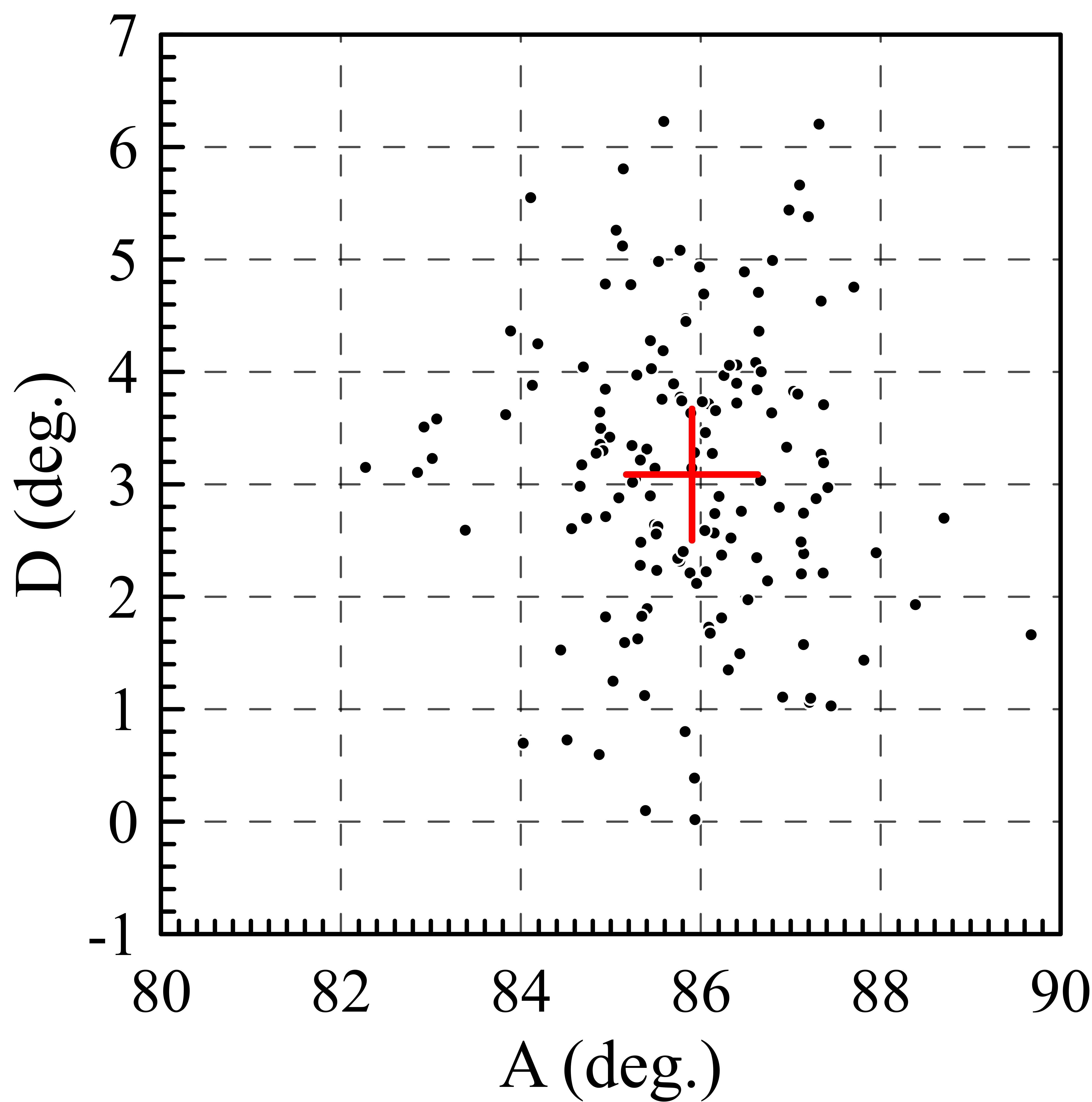}
\caption{The AD-diagrams for NGC 6793, with a cross mark indicating the position of the apex point ($A_{\rm o},~D_{\rm o}$).}
\label{fig:9}
\end{center}
\end{figure}

The apex coordinates ($A_{\rm o}$, $D_{\rm o}$) represent the convergent point toward which the stellar motions within the cluster are directed. This location corresponds to the intersection of the spatial velocity vectors of cluster members with the celestial sphere. \citet{Chupina_2001, Chupina_2006} proposed a method for calculating apex components in equatorial coordinates, using the mean space velocity vectors derived from Equation (\ref{eq:Vs}). These apex coordinates can be determined using the following approach:
\begin{equation}
\label{eq:ad}
\begin{split}
A_{\rm o} &= \tan^{-1} \Bigg(\frac{\overline{V_y}}{\overline{V_x}}\Bigg) \\
D_{\rm o} &= \tan^{-1} \Bigg(\frac{\overline{V_z}}{\sqrt {\overline{V_x^2} + \overline{V_{y}^{2}}}}\Bigg)
\end{split}
\end{equation}
The cross mark in Figure \ref{fig:9} illustrates the apex position for NGC 6793 and their coordinates in Table \ref{tab:VEP}.

\begin{table}
\caption{Calculated dynamical evolution and kinematic parameters for NGC 6793.}
\centering
\begin{tabular}{ll}
\hline
Parameter  &	Value	\\   \hline
$\overline{V}_X$ (km s$^{-1}$) & 1.82 $\pm$ 0.01	\\
$\overline{V}_Y$ (km s$^{-1}$) & 25.43 $\pm$ 5.04	\\
$\overline{V}_Z$ (km s$^{-1}$) & 1.42 $\pm$ 0.01	 \\
$A_{\rm o}$	($\deg$) &85.85 $\pm$ 0.11\\
$D_{\rm o}$	($\deg$) &3.12 $\pm$ 0 .57\\
$\overline{U}$ (km s$^{-1}$)&-22.94 $\pm$ 0.08	\\
$\overline{V}$ (km s$^{-1}$)&-9.44 $\pm$ 0.91	\\
$\overline{W}$ (km s$^{-1}$)	&-5.90 $\pm$ 0.14	\\
\hline
\label{tab:VEP}
\end{tabular}
\end{table}

\section{Galactic Orbital Parameters}
\label{sec_7}
The examination of kinematics and orbital dynamics provides crucial insights into the birth locations and Galactic distribution of open clusters \citep{Yontan_2023, Elsanhoury_2024}. For this study, we analyzed the orbit of NGC 6793 by integrating its motion within the Galactic potential. The analysis used the Python-based {\sc galpy} package \citep{Bovy_2015}\footnote{\url{https://galpy.readthedocs.io/en/v1.5.0/}}, utilising the {\sc MWPotential2014} model. This model combines three components: a spherical bulge \citep{Bovy_2015}, a disc described by \citet{Miyamoto_1975}, and a spherical dark matter halo based on \citet{Navarro_1996}. For the calculations, the Galactocentric distance of the Sun was assumed to be $R_{\rm gc}=8$ kpc, with a rotational velocity of $V_{\rm rot}=220$ km s$^{-1}$ \citep{Bovy_2012, Bovy_2015}. 

The Sun's distance from the Galactic plane was adopted as $25 \pm 5$ pc \citep{Juric_2008}. Radial velocity measurements are essential for orbit integration, and we used \textit{Gaia} DR3 radial velocity data to calculate the mean radial velocity and its uncertainty for NGC 6793. Stars with high membership probabilities ($P \geq 0.5$) were considered, resulting in 58 probable members. A weighted mean approach was applied \citep[see equations in][]{Soubiran_2018}, resulting in a mean radial velocity of $V_{\rm R} = -20.87 \pm 0.67$ km s$^{-1}$. Comparisons with literature values (Table~\ref{tab:Table-1}) include $-16.59 \pm 1.45$ km s$^{-1}$ \citep{Soubiran_2018}, $-25.68 \pm 4.71$ km s$^{-1}$ \citep{Dias_2021}, $-37 \pm 5.35$km s$^{-1}$ \citep{Tarricq_2022} and $-20.02 \pm 0.76$ km s$^{-1}$ \citep{Hunt_2024} are in a good agreement. Parameters used for orbit integration include equatorial coordinates ($\alpha=19^{\rm h} 23^{\rm m} 16^{\rm s}.08, \delta= +22^{\circ} 09^{\rm '} 32^{\rm''}.3$) from \citet{Cantat-Gaudin_2020}, proper motions ($\mu_{\alpha}\cos \delta$, $\mu_{\delta}$ = 3.814$\pm$0.031,~3.547$\pm$0.034 mas yr$^{-1}$) derived in Section~3.2, an isochrone distance ($d_{\rm iso} = 585 \pm 19$ pc) from Section~3.3 and presented in Table~\ref{tab:Final_table}.

\begin{figure*}
\centering
\includegraphics[width=1\textwidth]{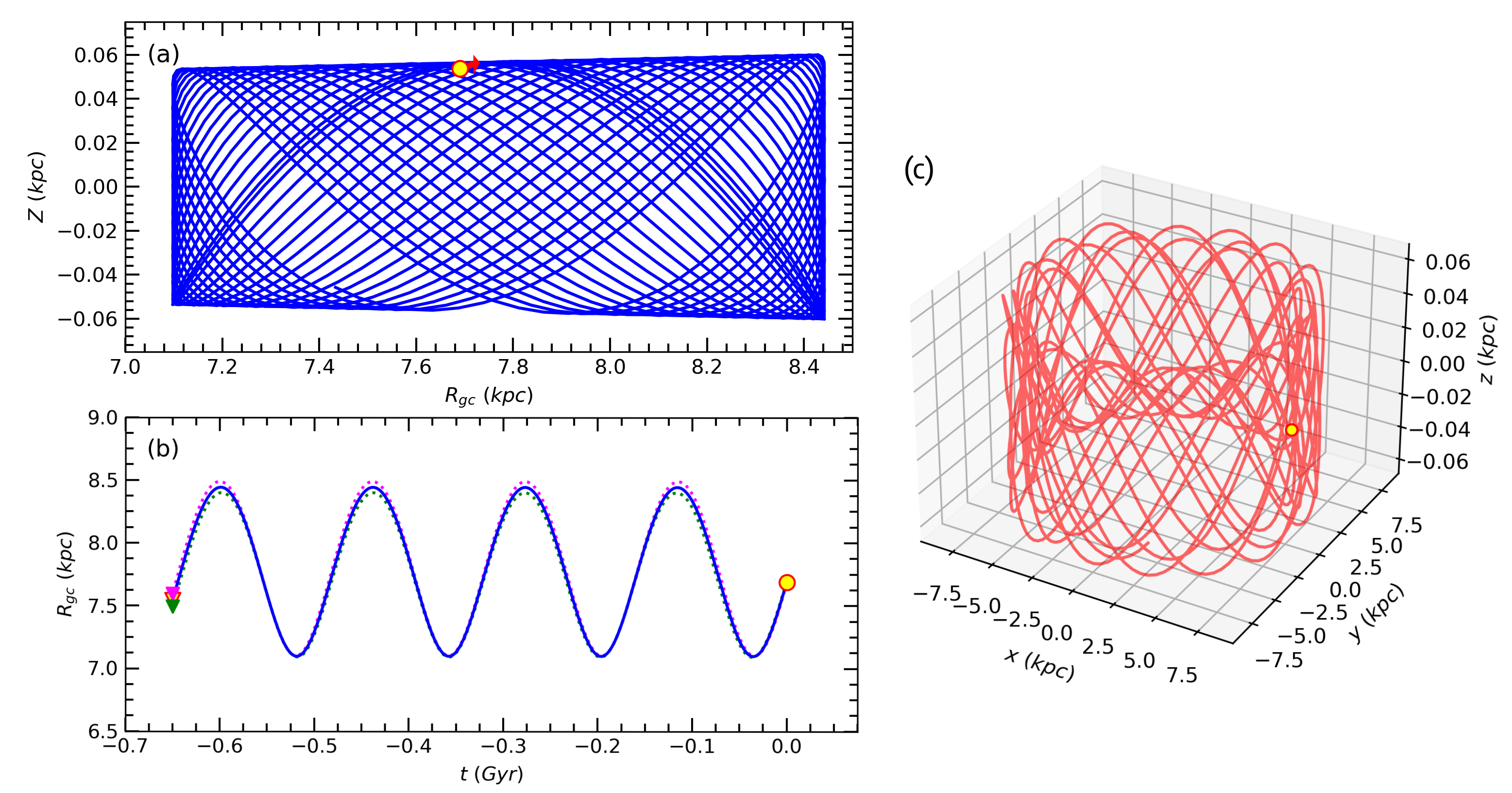}
\small
\caption{The Galactic trajectories and inferred birth radii of NGC 6793 are presented across three planes: $Z\times R_{\rm gc}$ (a), $R_{\rm gc}\times t$ (b), and $X$-$Y$-$Z \times t$ (c). Filled yellow circles mark the OC's current position, while triangles represent the estimated birth locations. The red arrow illustrates the direction of motion. Purple and pink dotted lines trace orbital paths considering uncertainties in the input parameters, with corresponding filled triangles denoting the lower and upper bounds of the birth position estimates.
\label{fig:Galactic_orbits}}
\end {figure*}

The orbit of NGC 6793 was integrated forward with a time step of 1 Myr over 2.5 Gyr to determine its likely current position. Figure~\ref{fig:Galactic_orbits}a depicts the cluster's path on the $Z \times R_{\rm gc}$ plane, showing its height above the Galactic plane and distance from the Galactic centre. To trace the cluster's origin, backward integration was performed for a duration matching its estimated age (650 Myr). Integration did not exceed the age of the cluster to reduce the potential inaccuracies arising from the limitations of the model and observational uncertainties, including distance, proper motion, and radial velocity errors \citep{Gaia_DR2, Sariya_2021}. Figure~\ref{fig:Galactic_orbits}b illustrates the time evolution of the cluster's Galactocentric distance on the $R_{\rm gc} \times t$ plane, accounting for input parameter variations. The results indicate an uncertainty of around 0.09 pc for the cluster's birth radius, suggesting that NGC 6793 formed outside the solar region, with an estimated birth radius of 7.54 kpc.

Orbit integration for NGC 6793 yielded the following orbital parameters: apogalactic ($R_{\rm a}=8442\pm 43$ pc) and perigalactic ($R_{\rm p}=7098\pm 5$ pc) distances, eccentricity ($e=0.087\pm 0.003$), maximum vertical distance from the Galactic plane ($Z_{\rm max}=60\pm 1$ pc) and orbital period ($P_{\rm orb}=217\pm 1$ Myr). The calculated apogalactic and perigalactic distances indicate that the orbit of NGC 6793 lies entirely beyond the solar circle (Figure~\ref{fig:Galactic_orbits}a). The maximum height above the Galactic plane, $Z_{\rm max} = 60\pm 1$ pc, further suggests that NGC 6793 is a member of the thin-disc population of the Milky Way \citep{Bilir_2006b, Bilir_2006c, Bilir_2008}.

\section{Conclusion}
\label{sec:Conclusion}
In this study, we conducted a comprehensive analysis of the open cluster NGC 6793, utilizing \textit{Gaia} DR3 data for photometric, astrometric, and kinematic assessments. A total of 147 stars were selected as the most probable members based on membership probabilities $P \geq 0.5$. These member stars served as the foundation for deriving key astrophysical and Galactic orbital parameters. The cluster's age, distance modulus, and reddening were simultaneously determined from the \textit{Gaia}-based CMD. A summary of the key findings is presented in Table~\ref{tab:Final_table}. The results from this study can be outlined as follows:

\begin{enumerate}
\item{We performed a detailed analysis of the radial distribution of stars in NGC 6793 to derive its structural parameters. The background stellar density, central density, and core radius were estimated to be $f_{\rm bg} = 62.855 \pm 0.750$ stars arcmin$^{-2}$, $f_{\rm 0} = 89.628 \pm 5.878$ stars arcmin$^{-2}$, and $r_{\rm c} = 1.593 \pm 0.128$ arcmin, respectively. The outermost boundary of the cluster was defined by the point where the radial density profile (RDP) aligns with the background stellar density. Based on this fitting, the limiting radius of the cluster was determined to be $r_{\rm lim}^{\rm obs} = 10$ arcmin.}

\item{Using the vector point diagram (VPD), we calculated the mean proper motion components for NGC 6793, obtaining as $(\mu_{\alpha}\cos\delta,~\mu_{\delta}) = (3.814 \pm 0.031,~3.547 \pm 0.034)$ mas yr$^{-1}$.}

\item{The colour excess for NGC 6793, derived from \textit{Gaia} data, was determined as $E(G_{\rm BP}-G_{\rm RP})=0.361 \pm 0.035$ mag by comparing the cluster's CMD with theoretical isochrones from \citet{Bressan_2012} at $z=0.0098$. Using the relation $E(G_{\rm BP}-G_{\rm RP}) = 1.41 \times E(B-V)$ from \citet{Sun_2021}, we estimated the $U\!BV$-system colour excess as $E(B-V) = 0.256 \pm 0.025$ mag.}

\item{The distance to NGC 6793, derived from isochrone fitting, was found to be $d_{\rm iso} = 585 \pm 19$ pc, which closely corresponds to the distance calculated from the trigonometric parallax, $d_{\varpi} = 597 \pm 26$ pc. The age of the cluster was estimated to be $t = 650 \pm 50$ Myr, based on the comparison between the observed CMD and the isochrone model from \citet{Bressan_2012} with $z = 0.0098$.}

\item{The mass function slope of NGC 6793 was determined to be $\Gamma = 1.40 \pm 0.26$. This value is in close agreement with the classic Salpeter value of 2.35 \citep{Salpeter_1955}. The total mass of the cluster was estimated to be approximately $139\pm12~M/M_{\odot}$. \citet{Yontan_2019} derived the slope of the mass function for NGC 6793 as $\Gamma = 1.31 \pm 0.31$ which is in good agreement with our findings}.

\item{Evolving times determination of the NGC 6793 OC reveals that the cluster is dynamically relaxed since the dynamical evolution parameter $\tau \gg 1$.}

\item{Equatorial coordinates of the coherent convergent point is computed with well known AD-digram method and numerically have values of ($A_{\rm o},~D_{\rm o}$)~=~($85^{\rm o}.85\pm0^{\rm o}.11,~3^{\rm o}.12\pm0^{\rm o}.57$).}

\item{Galactic orbit analysis revealed that NGC 6793 follows an elliptical orbit outside the solar radius, situated within the young thin-disc population of the Milky Way. Additionally, the estimated birth radius of the cluster, $\sim 7.54 \pm 0.01$ kpc, suggests that it originated beyond the solar neighbourhood.}
\end{enumerate}

\begin{table}
\renewcommand{\arraystretch}{1.2}
\setlength{\tabcolsep}{7pt}
  \centering
  \caption{Fundamental parameters of NGC 6793.}
  {\normalsize
        \begin{tabular}{lr}
\hline
Parameter & Value\\
\hline
($\alpha,~\delta)_{\rm J2000}$ & 19$^{\rm h}$23$^{\rm m}$16$^{\rm s}$.08, $+$22$^{\rm o}$09${\rm '}$32${\rm ''}$.3 \\
($l, b)_{\rm J2000}$                        & 56.18600, $+3.31804$    \\    
$f_{0}$ (stars arcmin$^{-2}$)               &  89.628$\pm$5.878          \\
$f_{\rm bg}$ (stars arcmin$^{-2}$)          & 62.855$\pm$0.750           \\
$r_{\rm c}$ (arcmin)                        & 1.593$\pm$0.128           \\
$r_{\rm c}$ (pc)                            & 0.28$\pm$0.01\\
$r_{\rm lim}$ (arcmin)                      & 10   \\
$r_{\rm lim}$ (pc)                          & 1.70  \\
$r_{\rm t}$ (pc)                            & 7.11$\pm$0.38\\
$R_{\rm h}$ (pc)                            & 0.773$\pm$0.01\\
Cluster members ($P\geq0.5$)                & 147                   \\
$\mu_{\alpha}\cos \delta$ (mas yr$^{-1}$)   & $3.814\pm 0.031$       \\
$\mu_{\delta}$ (mas yr$^{-1}$)              & $3.547 \pm 0.034$      \\
$\varpi$ (mas)                              & $1.674 \pm 0.045$       \\
$d_{\varpi}$ (pc)                           & $597\pm 26$             \\
$E(B-V)$ (mag)                              & $0.256\pm 0.025$ \\
$E(G_{\rm BP}-G_{\rm RP})$ (mag)            & $0.361\pm 0.035$ \\
$A_{\rm G}$ (mag)                           & $0.672\pm 0.065$ \\
$[{\rm Fe/H}]$ (dex)                        & $0.10 \pm 0.05$ \\
Age (Myr)                                   & $650 \pm 50$            \\
Distance modulus (mag)                      & $9.508 \pm0.070$ \\
Isochrone distance (pc)                     & $585 \pm19$              \\
$(X, Y, Z)_{\odot}$ (pc)                    & ($325\pm18, 485\pm22, 34\pm6$)\\
$R_{\rm gc}$ (kpc)                          & $7.69 \pm 0.01$           \\
MF slope                                    & 1.40 $\pm$ 0.26          \\
Total mass ($M_\odot$)                      & 139 $\pm$ 12          \\
$V_{\rm R}$ (km s$^{-1}$)                   & -20.87 $\pm$ 0.67         \\
$U_{\rm LSR}$ (km s$^{-1}$)                 & $-14.11\pm 0.25$         \\
$V_{\rm LSR}$ (kms$^{-1}$)                  & $4.75 \pm 0.97$         \\
$W_{\rm LSR}$ (kms$^{-1}$)                  & $0.67\pm0.26$          \\
$S_{_{\rm LSR}}$ (kms$^{-1}$)               & $14.90\pm1.04$          \\
$R_{\rm a}$ (pc)                            & $8442\pm 43$           \\
$R_{\rm p}$ (pc)                            & $7098 \pm 5.00$           \\
$z_{\rm max}$ (pc)                          & $60\pm 1.00$             \\
$e$                                         & $0.087\pm 0.003$        \\
$P_{\rm orb}$ (Myr)                         & $217 \pm 1.00$            \\
Birthplace (pc)                             & $7542 \pm 99.00$         \\
\hline
        \end{tabular}%
    } 
    \label{tab:Final_table}%
\end{table}%

\medskip

\section*{Acknowledgements}
 This study has been supported in part by the Scientific and Technological Research Council (T\"UB\.ITAK) 122F109. This research has made use of the WEBDA database, operated at the Department of Theoretical Physics and Astrophysics of the Masaryk University. We also made use of NASA's Astrophysics Data System as well as the VizieR and Simbad databases at CDS, Strasbourg, France and data from the European Space Agency (ESA) mission \emph{Gaia}\footnote{https://www.cosmos.esa.int/gaia}, processed by the \emph{Gaia} Data Processing and Analysis Consortium (DPAC)\footnote{https://www.cosmos.esa.int/web/gaia/dpac/consortium}. Funding for DPAC has been provided by national institutions, in particular, the institutions participating in the \emph{Gaia} Multilateral Agreement.



\bibliographystyle{mnras}
\bibliography{refs}

\begin{thebibliography}{}
\makeatletter
\relax
\def\mn@urlcharsother{\let\do\@makeother \do\$\do\&\do\#\do\^\do\_\do\%\do\~}
\def\mn@doi{\begingroup\mn@urlcharsother \@ifnextchar [ {\mn@doi@} {\mn@doi@[]}}
\def\mn@doi@[#1]#2{\def\@tempa{#1}\ifx\@tempa\@empty \href {http://dx.doi.org/#2} {doi:#2}\else \href {http://dx.doi.org/#2} {#1}\fi \endgroup}
\def\mn@eprint#1#2{\mn@eprint@#1:#2::\@nil}
\def\mn@eprint@arXiv#1{\href {http://arxiv.org/abs/#1} {{\tt arXiv:#1}}}
\def\mn@eprint@dblp#1{\href {http://dblp.uni-trier.de/rec/bibtex/#1.xml} {dblp:#1}}
\def\mn@eprint@#1:#2:#3:#4\@nil{\def\@tempa {#1}\def\@tempb {#2}\def\@tempc {#3}\ifx \@tempc \@empty \let \@tempc \@tempb \let \@tempb \@tempa \fi \ifx \@tempb \@empty \def\@tempb {arXiv}\fi \@ifundefined {mn@eprint@\@tempb}{\@tempb:\@tempc}{\expandafter \expandafter \csname mn@eprint@\@tempb\endcsname \expandafter{\@tempc}}}

\bibitem[\protect\citeauthoryear{{Adams} \& {Myers}}{{Adams} \& {Myers}}{2001}]{Adams_2001}
{Adams} F.~C.,  {Myers} P.~C.,  2001, \mn@doi [\apj] {10.1086/320941}, \href {https://ui.adsabs.harvard.edu/abs/2001ApJ...553..744A} {553, 744}

\bibitem[\protect\citeauthoryear{{Bastian}, {Covey}  \& {Meyer}}{{Bastian} et~al.}{2010}]{Bastian_2010}
{Bastian} N.,  {Covey} K.~R.,   {Meyer} M.~R.,  2010, \mn@doi [\araa] {10.1146/annurev-astro-082708-101642}, \href {https://ui.adsabs.harvard.edu/abs/2010ARA&A..48..339B} {48, 339}

\bibitem[\protect\citeauthoryear{{Baumgardt} \& {Makino}}{{Baumgardt} \& {Makino}}{2003}]{Baumgardt_2003}
{Baumgardt} H.,  {Makino} J.,  2003, \mn@doi [\mnras] {10.1046/j.1365-8711.2003.06286.x}, \href {https://ui.adsabs.harvard.edu/abs/2003MNRAS.340..227B} {340, 227}

\bibitem[\protect\citeauthoryear{{Bilir}, {Karaali}, {Ak}, {Yaz}  \& {Hamzao{\u{g}}lu}}{{Bilir} et~al.}{2006a}]{Bilir_2006b}
{Bilir} S.,  {Karaali} S.,  {Ak} S.,  {Yaz} E.,   {Hamzao{\u{g}}lu} E.,  2006a, \mn@doi [\na] {10.1016/j.newast.2006.10.001}, 12, 234

\bibitem[\protect\citeauthoryear{{Bilir}, {Karaali}  \& {Gilmore}}{{Bilir} et~al.}{2006b}]{Bilir_2006c}
{Bilir} S.,  {Karaali} S.,   {Gilmore} G.,  2006b, \mn@doi [\mnras] {10.1111/j.1365-2966.2006.09891.x}, 366, 1295

\bibitem[\protect\citeauthoryear{{Bilir}, {Cabrera-Lavers}, {Karaali}, {Ak}, {Yaz}  \& {L{\'o}pez-Corredoira}}{{Bilir} et~al.}{2008}]{Bilir_2008}
{Bilir} S.,  {Cabrera-Lavers} A.,  {Karaali} S.,  {Ak} S.,  {Yaz} E.,   {L{\'o}pez-Corredoira} M.,  2008, \mn@doi [\pasa] {10.1071/AS07026}, 25, 69

\bibitem[\protect\citeauthoryear{{Binney} \& {Merrifield}}{{Binney} \& {Merrifield}}{1998}]{Binney_1998}
{Binney} J.,  {Merrifield} M.,  1998, {Galactic Astronomy}

\bibitem[\protect\citeauthoryear{Binney \& Tremaine}{Binney \& Tremaine}{2008}]{Binney_Tremaine_2008}
Binney J.,  Tremaine S.,  2008, Galactic Dynamics, 2nd edn.
Princeton University Press

\bibitem[\protect\citeauthoryear{{Bisht}, {Yadav}, {Ganesh}, {Durgapal}, {Rangwal}  \& {Fynbo}}{{Bisht} et~al.}{2019}]{Bisht_2019}
{Bisht} D.,  {Yadav} R.~K.~S.,  {Ganesh} S.,  {Durgapal} A.~K.,  {Rangwal} G.,   {Fynbo} J.~P.~U.,  2019, \mn@doi [\mnras] {10.1093/mnras/sty2781}, \href {https://ui.adsabs.harvard.edu/abs/2019MNRAS.482.1471B} {482, 1471}

\bibitem[\protect\citeauthoryear{{Bisht}, {Zhu}, {Yadav}, {Durgapal}  \& {Rangwal}}{{Bisht} et~al.}{2020}]{Bisht_2020}
{Bisht} D.,  {Zhu} Q.,  {Yadav} R.~K.~S.,  {Durgapal} A.,   {Rangwal} G.,  2020, \mn@doi [\mnras] {10.1093/mnras/staa656}, 494, 607

\bibitem[\protect\citeauthoryear{{Bossini} et~al.,}{{Bossini} et~al.}{2019}]{Bossini_2019}
{Bossini} D.,  et~al., 2019, \mn@doi [\aap] {10.1051/0004-6361/201834693}, \href {https://ui.adsabs.harvard.edu/abs/2019A&A...623A.108B} {623, A108}

\bibitem[\protect\citeauthoryear{{Bovy}}{{Bovy}}{2015}]{Bovy_2015}
{Bovy} J.,  2015, \mn@doi [\apjs] {10.1088/0067-0049/216/2/29}, \href {https://ui.adsabs.harvard.edu/abs/2015ApJS..216...29B} {216, 29}

\bibitem[\protect\citeauthoryear{{Bovy} \& {Tremaine}}{{Bovy} \& {Tremaine}}{2012}]{Bovy_2012}
{Bovy} J.,  {Tremaine} S.,  2012, \mn@doi [\apj] {10.1088/0004-637X/756/1/89}, \href {https://ui.adsabs.harvard.edu/abs/2012ApJ...756...89B} {756, 89}

\bibitem[\protect\citeauthoryear{Bressan, Marigo, Girardi, Salasnich, Dal~Cero, Rubele  \& Nanni}{Bressan et~al.}{2012}]{Bressan_2012}
Bressan A.,  Marigo P.,  Girardi L.,  Salasnich B.,  Dal~Cero C.,  Rubele S.,   Nanni A.,  2012, \mn@doi [\mnras] {10.1111/j.1365-2966.2012.21948.x}, 427, 127

\bibitem[\protect\citeauthoryear{{Bukowiecki}, {Maciejewski}, {Konorski}  \& {Strobel}}{{Bukowiecki} et~al.}{2011}]{Bukowiecki_2011}
{Bukowiecki} {\L}.,  {Maciejewski} G.,  {Konorski} P.,   {Strobel} A.,  2011, \mn@doi [\actaa] {10.48550/arXiv.1107.5119}, 61, 231

\bibitem[\protect\citeauthoryear{Cantat-Gaudin \& Anders}{Cantat-Gaudin \& Anders}{2020}]{Cantat-Gaudin-Anders_2020}
Cantat-Gaudin T.,  Anders F.,  2020, \mn@doi [\aap] {10.1051/0004-6361/201936691}, 633, A99

\bibitem[\protect\citeauthoryear{Cantat-Gaudin et~al.,}{Cantat-Gaudin et~al.}{2020}]{Cantat-Gaudin_2020}
Cantat-Gaudin T.,  et~al., 2020, \mn@doi [\aap] {10.1051/0004-6361/202038192}, 640, A1

\bibitem[\protect\citeauthoryear{{Carraro}, {Sales Silva}, {Moni Bidin}  \& {Vazquez}}{{Carraro} et~al.}{2017}]{Carraro_2017}
{Carraro} G.,  {Sales Silva} J.~V.,  {Moni Bidin} C.,   {Vazquez} R.~A.,  2017, \mn@doi [\aj] {10.3847/1538-3881/153/3/99}, 153, 99

\bibitem[\protect\citeauthoryear{{Chupina}, {Reva}  \& {Vereshchagin}}{{Chupina} et~al.}{2001}]{Chupina_2001}
{Chupina} N.~V.,  {Reva} V.~G.,   {Vereshchagin} S.~V.,  2001, \mn@doi [\aap] {10.1051/0004-6361:20010337}, \href {https://ui.adsabs.harvard.edu/abs/2001A&A...371..115C} {371, 115}

\bibitem[\protect\citeauthoryear{{Chupina}, {Reva}  \& {Vereshchagin}}{{Chupina} et~al.}{2006}]{Chupina_2006}
{Chupina} N.~V.,  {Reva} V.~G.,   {Vereshchagin} S.~V.,  2006, \mn@doi [\aap] {10.1051/0004-6361:20054009}, \href {https://ui.adsabs.harvard.edu/abs/2006A&A...451..909C} {451, 909}

\bibitem[\protect\citeauthoryear{Co{\c{s}}kuno{\v{g}}lu et~al.,}{Co{\c{s}}kuno{\v{g}}lu et~al.}{2011}]{Coskunoglu_2011}
Co{\c{s}}kuno{\v{g}}lu B.,  et~al., 2011, \mn@doi [\mnras] {10.1111/j.1365-2966.2010.17983.x}, 412, 1237

\bibitem[\protect\citeauthoryear{Dias, Monteir, Moitinho, L{\'e}pine, Carraro, Paunzen, Alessi  \& Villela}{Dias et~al.}{2021}]{Dias_2021}
Dias W.~S.,  Monteir H.,  Moitinho A.,  L{\'e}pine J.~R.~D.,  Carraro G.,  Paunzen E.,  Alessi B.,   Villela L.,  2021, \mn@doi [\mnras] {10.1093/mnras/stab770}, 504, 356

\bibitem[\protect\citeauthoryear{{Dib} \& {Basu}}{{Dib} \& {Basu}}{2018}]{Dib_2018}
{Dib} S.,  {Basu} S.,  2018, \mn@doi [\aap] {10.1051/0004-6361/201732490}, \href {https://ui.adsabs.harvard.edu/abs/2018A&A...614A..43D} {614, A43}

\bibitem[\protect\citeauthoryear{{Donada} et~al.,}{{Donada} et~al.}{2023}]{Donada_2023}
{Donada} J.,  et~al., 2023, \mn@doi [\aap] {10.1051/0004-6361/202245219}, \href {https://ui.adsabs.harvard.edu/abs/2023A&A...675A..89D} {675, A89}

\bibitem[\protect\citeauthoryear{{Elsanhoury}, {Postnikova}, {Chupina}, {Vereshchagin}, {Sariya}, {Yadav}  \& {Jiang}}{{Elsanhoury} et~al.}{2018}]{Elsanhoury_2018}
{Elsanhoury} W.~H.,  {Postnikova} E.~S.,  {Chupina} N.~V.,  {Vereshchagin} S.~V.,  {Sariya} D.~P.,  {Yadav} R.~K.~S.,   {Jiang} I.-G.,  2018, \mn@doi [\apss] {10.1007/s10509-018-3268-3}, \href {https://ui.adsabs.harvard.edu/abs/2018Ap&SS.363...58E} {363, 58}

\bibitem[\protect\citeauthoryear{{Elsanhoury}, {A}, {Elkholy}  \& {{\c{C}}{\i}nar}}{{Elsanhoury} et~al.}{2024}]{Elsanhoury_2024}
{Elsanhoury} W.~H.,  {A} H.~A.,  {Elkholy} E.~A.,   {{\c{C}}{\i}nar} D.~C.,  2024, \mn@doi [arXiv e-prints] {10.48550/arXiv.2412.07871}, \href {https://ui.adsabs.harvard.edu/abs/2024arXiv241207871E} {p. arXiv:2412.07871}

\bibitem[\protect\citeauthoryear{{Evans} et~al.,}{{Evans} et~al.}{2018}]{Evans_2018}
{Evans} D.~W.,  et~al., 2018, \mn@doi [\aap] {10.1051/0004-6361/201832756}, \href {https://ui.adsabs.harvard.edu/abs/2018A&A...616A...4E} {616, A4}

\bibitem[\protect\citeauthoryear{{Fich} \& {Tremaine}}{{Fich} \& {Tremaine}}{1991}]{Fich_1991}
{Fich} M.,  {Tremaine} S.,  1991, \mn@doi [\araa] {10.1146/annurev.aa.29.090191.002205}, \href {https://ui.adsabs.harvard.edu/abs/1991ARA&A..29..409F} {29, 409}

\bibitem[\protect\citeauthoryear{{Fukushige} \& {Heggie}}{{Fukushige} \& {Heggie}}{2000}]{Fukushige_2000}
{Fukushige} T.,  {Heggie} D.~C.,  2000, \mn@doi [\mnras] {10.1046/j.1365-8711.2000.03811.x}, \href {https://ui.adsabs.harvard.edu/abs/2000MNRAS.318..753F} {318, 753}

\bibitem[\protect\citeauthoryear{{Gaia Collaboration} et~al.,}{{Gaia Collaboration} et~al.}{2018}]{Gaia_DR2}
{Gaia Collaboration} et~al., 2018, \mn@doi [\aap] {10.1051/0004-6361/201833051}, \href {https://ui.adsabs.harvard.edu/abs/2018A&A...616A...1G} {616, A1}

\bibitem[\protect\citeauthoryear{{Gaia Collaboration} et~al.,}{{Gaia Collaboration} et~al.}{2021}]{Gaia_EDR3}
{Gaia Collaboration} et~al., 2021, \mn@doi [\aap] {10.1051/0004-6361/202039657}, \href {https://ui.adsabs.harvard.edu/abs/2021A&A...649A...1G} {649, A1}

\bibitem[\protect\citeauthoryear{{Gaia Collaboration} et~al.,}{{Gaia Collaboration} et~al.}{2023}]{Gaia_DR3}
{Gaia Collaboration} et~al., 2023, \mn@doi [\aap] {10.1051/0004-6361/202243940}, \href {https://ui.adsabs.harvard.edu/abs/2023A&A...674A...1G} {674, A1}

\bibitem[\protect\citeauthoryear{{Gokmen}, {Eker}, {Yontan}, {Bilir}, {Ak}, {Ak}, {Banks}  \& {Sarajedini}}{{Gokmen} et~al.}{2023}]{Gökmen_2023}
{Gokmen} S.,  {Eker} Z.,  {Yontan} T.,  {Bilir} S.,  {Ak} T.,  {Ak} S.,  {Banks} T.,   {Sarajedini} A.,  2023, \mn@doi [\aj] {10.3847/1538-3881/ad08b0}, \href {https://ui.adsabs.harvard.edu/abs/2023AJ....166..263G} {166, 263}

\bibitem[\protect\citeauthoryear{{Hunt} \& {Reffert}}{{Hunt} \& {Reffert}}{2024}]{Hunt_2024}
{Hunt} E.~L.,  {Reffert} S.,  2024, \mn@doi [\aap] {10.1051/0004-6361/202348662}, \href {https://ui.adsabs.harvard.edu/abs/2024A&A...686A..42H} {686, A42}

\bibitem[\protect\citeauthoryear{{Inagaki} \& {Saslaw}}{{Inagaki} \& {Saslaw}}{1985}]{Inagaki_1985}
{Inagaki} S.,  {Saslaw} W.~C.,  1985, \mn@doi [\apj] {10.1086/163164}, \href {https://ui.adsabs.harvard.edu/abs/1985ApJ...292..339I} {292, 339}

\bibitem[\protect\citeauthoryear{Joshi, Dambis, Pandey  \& Joshi}{Joshi et~al.}{2016}]{Joshi_2016}
Joshi Y.~C.,  Dambis A.~K.,  Pandey A.~K.,   Joshi S.,  2016, \mn@doi [\aap] {10.1051/0004-6361/201628944}, 593, A116

\bibitem[\protect\citeauthoryear{{Juri{\'c}} et~al.,}{{Juri{\'c}} et~al.}{2008}]{Juric_2008}
{Juri{\'c}} M.,  et~al., 2008, \mn@doi [\apj] {10.1086/523619}, \href {https://ui.adsabs.harvard.edu/abs/2008ApJ...673..864J} {673, 864}

\bibitem[\protect\citeauthoryear{King}{King}{1962}]{King_1962}
King I.,  1962, \mn@doi [\aj] {10.1086/108756}, 67, 471

\bibitem[\protect\citeauthoryear{Kroupa}{Kroupa}{2001}]{Kroupa_2001}
Kroupa P.,  2001, \mn@doi [\mnras] {10.1046/j.1365-8711.2001.04022.x}, 322, 231

\bibitem[\protect\citeauthoryear{{K{\"u}pper}, {MacLeod}  \& {Heggie}}{{K{\"u}pper} et~al.}{2008}]{Küpper_2008}
{K{\"u}pper} A. H.~W.,  {MacLeod} A.,   {Heggie} D.~C.,  2008, \mn@doi [\mnras] {10.1111/j.1365-2966.2008.13323.x}, \href {https://ui.adsabs.harvard.edu/abs/2008MNRAS.387.1248K} {387, 1248}

\bibitem[\protect\citeauthoryear{Lada \& Lada}{Lada \& Lada}{2003}]{Lada_2003}
Lada C.~J.,  Lada E.~A.,  2003, \mn@doi [\araa] {10.1146/annurev.astro.41.011802.094844}, 41, 57

\bibitem[\protect\citeauthoryear{Leggett}{Leggett}{1992}]{Leggett_1992}
Leggett S.~K.,  1992, \mn@doi [\apjs] {10.1086/191720}, 82, 351

\bibitem[\protect\citeauthoryear{Liu \& Pang}{Liu \& Pang}{2019}]{Liu_2019}
Liu L.,  Pang X.,  2019, \mn@doi [\apjs] {10.3847/1538-4365/ab530a}, 245, 32

\bibitem[\protect\citeauthoryear{{Liu}, {Zhu}  \& {Hu}}{{Liu} et~al.}{2011}]{Liu11}
{Liu} J.~C.,  {Zhu} Z.,   {Hu} B.,  2011, \mn@doi [\aap] {10.1051/0004-6361/201116947}, \href {https://ui.adsabs.harvard.edu/abs/2011A&A...536A.102L} {536, A102}

\bibitem[\protect\citeauthoryear{{Maciejewski} \& {Niedzielski}}{{Maciejewski} \& {Niedzielski}}{2007}]{Maciejewski_2007}
{Maciejewski} G.,  {Niedzielski} A.,  2007, \mn@doi [\aap] {10.1051/0004-6361:20066588}, \href {https://ui.adsabs.harvard.edu/abs/2007A&A...467.1065M} {467, 1065}

\bibitem[\protect\citeauthoryear{{Mathieu} \& {Latham}}{{Mathieu} \& {Latham}}{1986}]{Mathieu_1986}
{Mathieu} R.~D.,  {Latham} D.~W.,  1986, \mn@doi [\aj] {10.1086/114269}, \href {https://ui.adsabs.harvard.edu/abs/1986AJ.....92.1364M} {92, 1364}

\bibitem[\protect\citeauthoryear{{Melchior}}{{Melchior}}{1958}]{Melchior1958}
{Melchior} P.~J.,  1958, Ciel et Terre, \href {https://ui.adsabs.harvard.edu/abs/1958C&T....74R.290M} {74, 290}

\bibitem[\protect\citeauthoryear{{Miyamoto} \& {Nagai}}{{Miyamoto} \& {Nagai}}{1975}]{Miyamoto_1975}
{Miyamoto} M.,  {Nagai} R.,  1975, \pasj, \href {https://ui.adsabs.harvard.edu/abs/1975PASJ...27..533M} {27, 533}

\bibitem[\protect\citeauthoryear{{Navarro}, {Frenk}  \& {White}}{{Navarro} et~al.}{1996}]{Navarro_1996}
{Navarro} J.~F.,  {Frenk} C.~S.,   {White} S. D.~M.,  1996, \mn@doi [\apj] {10.1086/177173}, \href {https://ui.adsabs.harvard.edu/abs/1996ApJ...462..563N} {462, 563}

\bibitem[\protect\citeauthoryear{{Perryman} et~al.,}{{Perryman} et~al.}{1998}]{Perryman_1998}
{Perryman} M.~A.~C.,  et~al., 1998, \mn@doi [\aap] {10.48550/arXiv.astro-ph/9707253}, \href {https://ui.adsabs.harvard.edu/abs/1998A&A...331...81P} {331, 81}

\bibitem[\protect\citeauthoryear{{Phelps} \& {Janes}}{{Phelps} \& {Janes}}{1993}]{Phelps_janes_1993}
{Phelps} R.~L.,  {Janes} K.~A.,  1993, \mn@doi [\aj] {10.1086/116772}, \href {https://ui.adsabs.harvard.edu/abs/1993AJ....106.1870P} {106, 1870}

\bibitem[\protect\citeauthoryear{{Piskunov}, {Belikov}, {Kharchenko}, {Sagar}  \& {Subramaniam}}{{Piskunov} et~al.}{2004}]{Piskunov_2004}
{Piskunov} A.~E.,  {Belikov} A.~N.,  {Kharchenko} N.~V.,  {Sagar} R.,   {Subramaniam} A.,  2004, \mn@doi [\mnras] {10.1111/j.1365-2966.2004.07620.x}, \href {https://ui.adsabs.harvard.edu/abs/2004MNRAS.349.1449P} {349, 1449}

\bibitem[\protect\citeauthoryear{{Poggio} et~al.,}{{Poggio} et~al.}{2021}]{Poggio_2021}
{Poggio} E.,  et~al., 2021, \mn@doi [\aap] {10.1051/0004-6361/202140687}, \href {https://ui.adsabs.harvard.edu/abs/2021A&A...651A.104P} {651, A104}

\bibitem[\protect\citeauthoryear{{Riello} et~al.,}{{Riello} et~al.}{2021}]{Riello_2021}
{Riello} M.,  et~al., 2021, \mn@doi [\aap] {10.1051/0004-6361/202039587}, 649, A3

\bibitem[\protect\citeauthoryear{{R{\"o}ser} \& {Schilbach}}{{R{\"o}ser} \& {Schilbach}}{2019}]{Röser_2019}
{R{\"o}ser} S.,  {Schilbach} E.,  2019, \mn@doi [\aap] {10.1051/0004-6361/201935502}, \href {https://ui.adsabs.harvard.edu/abs/2019A&A...627A...4R} {627, A4}

\bibitem[\protect\citeauthoryear{{Salpeter}}{{Salpeter}}{1955}]{Salpeter_1955}
{Salpeter} E.~E.,  1955, \mn@doi [\apj] {10.1086/145971}, 121, 161

\bibitem[\protect\citeauthoryear{{Sariya} et~al.,}{{Sariya} et~al.}{2021}]{Sariya_2021}
{Sariya} D.~P.,  et~al., 2021, \mn@doi [\aj] {10.3847/1538-3881/abd31d}, \href {https://ui.adsabs.harvard.edu/abs/2021AJ....161..101S} {161, 101}

\bibitem[\protect\citeauthoryear{Soubiran et~al.,}{Soubiran et~al.}{2018}]{Soubiran_2018}
Soubiran C.,  et~al., 2018, \mn@doi [\aap] {10.1051/0004-6361/201834020}, 619, A155

\bibitem[\protect\citeauthoryear{{Spitzer} \& {Hart}}{{Spitzer} \& {Hart}}{1971}]{Spitzer_1971}
{Spitzer} Jr. L.,  {Hart} M.~H.,  1971, \mn@doi [\apj] {10.1086/150855}, \href {https://ui.adsabs.harvard.edu/abs/1971ApJ...164..399S} {164, 399}

\bibitem[\protect\citeauthoryear{Sun, Jiang, Yuan  \& Li}{Sun et~al.}{2021}]{Sun_2021}
Sun M.,  Jiang B.,  Yuan H.,   Li J.,  2021, \mn@doi [\apjs] {10.3847/1538-4365/abf929}, 254, 38

\bibitem[\protect\citeauthoryear{{Tarricq} et~al.,}{{Tarricq} et~al.}{2021}]{Tarricq_2021}
{Tarricq} Y.,  et~al., 2021, \mn@doi [\aap] {10.1051/0004-6361/202039388}, \href {https://ui.adsabs.harvard.edu/abs/2021A&A...647A..19T} {647, A19}

\bibitem[\protect\citeauthoryear{{Tarricq}, {Soubiran}, {Casamiquela}, {Castro-Ginard}, {Olivares}, {Miret-Roig}  \& {Galli}}{{Tarricq} et~al.}{2022}]{Tarricq_2022}
{Tarricq} Y.,  {Soubiran} C.,  {Casamiquela} L.,  {Castro-Ginard} A.,  {Olivares} J.,  {Miret-Roig} N.,   {Galli} P.~A.~B.,  2022, \mn@doi [\aap] {10.1051/0004-6361/202142186}, \href {https://ui.adsabs.harvard.edu/abs/2022A&A...659A..59T} {659, A59}

\bibitem[\protect\citeauthoryear{{Yadav}, {Sariya}  \& {Sagar}}{{Yadav} et~al.}{2013}]{Yadav_2013}
{Yadav} R.~K.~S.,  {Sariya} D.~P.,   {Sagar} R.,  2013, \mn@doi [\mnras] {10.1093/mnras/stt136}, \href {https://ui.adsabs.harvard.edu/abs/2013MNRAS.430.3350Y} {430, 3350}

\bibitem[\protect\citeauthoryear{{Yontan}}{{Yontan}}{2023}]{Yontan_2023}
{Yontan} T.,  2023, \mn@doi [\aj] {10.3847/1538-3881/aca6f0}, 165, 79

\bibitem[\protect\citeauthoryear{Yontan et~al.,}{Yontan et~al.}{2019}]{Yontan_2019}
Yontan T.,  et~al., 2019, \mn@doi [Astrophysics and Space Science] {10.1007/s10509-019-3640-y}, 364, 152

\bibitem[\protect\citeauthoryear{{Yontan} et~al.,}{{Yontan} et~al.}{2022}]{Yontan_2022}
{Yontan} T.,  et~al., 2022, \mn@doi [\rmxaa] {10.22201/ia.01851101p.2022.58.02.14}, \href {https://ui.adsabs.harvard.edu/abs/2022RMxAA..58..333Y} {58, 333}

\bibitem[\protect\citeauthoryear{{Zhong}, {Chen}, {Wu}, {Li}, {Bai}  \& {Hou}}{{Zhong} et~al.}{2020}]{Zhong_2020}
{Zhong} J.,  {Chen} L.,  {Wu} D.,  {Li} L.,  {Bai} L.,   {Hou} J.,  2020, \mn@doi [\aap] {10.1051/0004-6361/201937131}, 640, A127

\bibitem[\protect\citeauthoryear{{{\v{S}}ablevi{\v{c}}i{\={u}}t{\.{e}}}, {Vansevi{\v{c}}ius}, {Kodaira}, {Narbutis}, {Stonkut{\.{e}}}  \& {Brid{\v{z}}ius}}{{{\v{S}}ablevi{\v{c}}i{\={u}}t{\.{e}}} et~al.}{2006}]{Sableviciute_2006}
{{\v{S}}ablevi{\v{c}}i{\={u}}t{\.{e}}} I.,  {Vansevi{\v{c}}ius} V.,  {Kodaira} K.,  {Narbutis} D.,  {Stonkut{\.{e}}} R.,   {Brid{\v{z}}ius} A.,  2006, \mn@doi [Baltic Astronomy] {10.48550/arXiv.astro-ph/0701774}, \href {https://ui.adsabs.harvard.edu/abs/2006BaltA..15..547S} {15, 547}

\makeatother
\end{thebibliography}

\bsp	
\label{lastpage}
\end{document}